\documentclass[aps,prx,twocolumn,superscriptaddress,showpacs,amsmath,amssymb,longbibliography, amsfonts,10pt]{revtex4-2}

\usepackage{amsmath}
\usepackage{amsfonts}
\usepackage{amssymb}
\usepackage{graphicx}
\usepackage{color}
\usepackage{xcolor}
\usepackage{bbold}
\usepackage{appendix}
\usepackage{placeins}
\usepackage[citecolor=blue,urlcolor=blue]{hyperref}
\hypersetup{colorlinks}
\usepackage{subfig}
\usepackage{bookmark}
\usepackage{makecell}
\usepackage{mathtools}
\usepackage{comment}
\usepackage{tikz}
\usetikzlibrary{arrows.meta}
\usepackage{array}
\usepackage{multirow}
\usepackage[nolist,nohyperlinks]{acronym}
\usepackage[export]{adjustbox}
\newacro{RG}[RG]{renomalization group}
\newacro{fdr}[FDR]{fluctuation-dissipation relation}
\newacro{EP}[EP]{exceptional point}
\usepackage[normalem]{ulem}
\usepackage[format=plain, justification=justified]{caption}

\newacro{CEP}[CEP]{critical exceptional point}
\newacro{frg}[FRG]{functional RG}
\newacro{DSE}[DSE]{Dyson-Schwinger equation}
\newacro{MSRJD}[MSRJD]{Martin-Siggia-Rose-Janssen-DeDominicis}
\newacro{BKT}[BKT]{Berezinskii–Kosterlitz–Thouless}
\newacro{KPZ}[KPZ]{Kardar–Parisi–Zhang}
\newacro{EFT}[EFT]{effective field theory}
\newacro{Nlsm}[NL$\sigma$M]{non-linear $\sigma$ model}
\usepackage{lipsum}

\newcommand{\dd}{\mathrm{d}}

\newcommand{\vecpi}{\boldsymbol\Pi}
\newcommand{\vecxi}{\boldsymbol\xi}
\newcommand{\vecphi}{\boldsymbol{\phi}}
\newcommand{\vecPhi}{\boldsymbol{\Phi}}
\newcommand{\vectheta}{\boldsymbol{\theta}}
\newcommand{\vecchi}{\boldsymbol{\chi}}
\newcommand{\vecr}{\boldsymbol{r}}
\newcommand{\vecj}{\boldsymbol{j}}
\newcommand{\vecbeta}{\boldsymbol{\beta}}

\newcommand{\vecpsi}{\boldsymbol{\psi}}
\newcommand{\vecvarphi}{\boldsymbol\varphi}

\newcommand{\vecq}{\mathbf{q}}
\newcommand{\vecp}{\mathbf{p}}

\definecolor{mygreen}{RGB}{0,160,60}

\begin{document}
\title{Far-from-equilibrium scaling of non-abelian Goldstone modes 
}
\author{Carl Philipp Zelle}
\affiliation{Department of Physics, Harvard University, Cambridge MA 02138, USA}
\author{Gustav John}
\affiliation{Institut f\"ur Theoretische Physik, Universit\"at zu K\"oln, 50937 Cologne, Germany}
\author{Orla Supple}
\affiliation{Institut f\"ur Theoretische Physik, Universit\"at zu K\"oln, 50937 Cologne, Germany}
\author{Romain Daviet}
\affiliation{Institut f\"ur Theoretische Physik, Universit\"at zu K\"oln, 50937 Cologne, Germany}
\author{Sebastian Diehl}
\affiliation{Institut f\"ur Theoretische Physik, Universit\"at zu K\"oln, 50937 Cologne, Germany}

\begin{abstract}
We identify a broad class of nonthermal phases generated by the interplay of continuous symmetry breaking and weak nonequilibrium driving. Extending Kardar–Parisi–Zhang (KPZ) physics beyond the single $SO(2)$ chronon associated with periodically broken time translations, we construct nonequilibrium nonlinear sigma models for $SO(2)\times O(N)$ symmetry, describing non-Abelian time crystals with coexisting temporal and internal order. This symmetry structure arises naturally in driven quantum materials, active matter, and optically induced periodic states. For rotating and oscillating phases, we derive the Goldstone theories and show that chronon–$O(N)$ couplings remain finite deep in the ordered regime. One-loop renormalization group analysis reveals a KPZ-like dimensional structure: in $d=1,2$, arbitrarily weak nonequilibrium perturbations destabilize the equilibrium fixed point and generate strongly coupled nonthermal fixed points, realizing emergent equilibrium breaking. By contrast, for $d\geq3$, weak perturbations are irrelevant and effective equilibrium is restored. A central result is unconventional weak dynamic scaling in the rotating phase: strongly coupled Goldstone sectors acquire distinct universal dynamical exponents despite belonging to the same order parameter. We characterize this scaling analytically and corroborate it through direct simulations in $1+1$ dimensions. In the oscillating phase, we recover and extend weak-scaling regimes known from drifting polymers. Finally, compactness and topological defects ultimately destroy long-range order but leave experimentally accessible nonthermal scaling windows. Together, these results extend KPZ universality to non-Abelian symmetry breaking.
\end{abstract}
\maketitle

\section{Introduction}

When does nonequilibrium driving matter? Experience from condensed matter physics suggests that thermodynamic equilibrium is remarkably robust: macroscopic properties typically remain unchanged even under weak external driving, such as shining light on a material. In other words, although the drive explicitly breaks equilibrium conditions microscopically, the macroscopic properties of the coarse-grained steady state and its relaxational dynamics are still effectively described by a free energy. Thermal equilibrium emerges at long wavelengths.

Yet the opposite can occur. Even weak nonequilibrium perturbations can drive a system into a phase with universal macroscopic scaling behavior that is impossible in equilibrium. In this case, the system realizes a genuinely \emph{nonthermal phase of matter}. The paradigmatic example is the Kardar-Parisi-Zhang (KPZ) equation \cite{Kardar1986}, originally introduced to describe the roughening of growing interfaces \cite{Krug1997}. In $1+1$ and $2+1$ dimensions, arbitrarily weak KPZ nonlinearities destabilize the diffusive fixed point and drive the system to a strongly coupled nonthermal regime with universal rough fluctuations. The resulting steady state exhibits nontrivial dynamical exponents, most notably $z=\frac{3}{2}$ in $1+1$D and $z\approx 1.6$ in $2+1$D \cite{Pagnani2015}. From a renormalization group perspective, the KPZ coupling is a relevant perturbation that universally drives the system toward a strongly interacting nonequilibrium fixed point. We refer to this phenomenon as emergent equilibrium breaking.

KPZ scaling has been observed experimentally in growing interfaces \cite{Takeuchi2010}, but its scope is much broader. More generally, KPZ behavior is expected in dynamical phases with time-dependent steady states \cite{Daviet2025}. Travelling-wave phases naturally arise in active matter \cite{Pisegna2024, Brauns2024} as well as in driven materials such as skyrmion lattices \cite{Tengdin22, delSer23, Rucker2026}. Recent theoretical work further indicates that photo-induced orders in optically pumped materials can occur at finite frequencies \cite{Zelle2026, okugawa2026, Diessel_2026, chattopadhyay2026, zelle2024}. In these systems, KPZ scaling no longer originates from linear interface growth, but from a form of ``periodic growth'' associated with the compact $SO(2)$ Goldstone mode of broken time translations in a time crystal~\cite{Daviet2025}. This mechanism has recently led to the first experimental observations of the universal KPZ scaling function in $1+1$~\cite{Fontaine2022} and $2+1$~\cite{Widmann2026} dimensional exciton-polariton condensates.

In this paper, we demonstrate that KPZ scaling is only the ``tip of the iceberg'' of a much broader nonequilibrium phenomenology. We identify novel universality classes of nonthermal phases of matter that generalize the KPZ scenario to systems with additional spontaneously broken continuous symmetries. This extension is motivated by the rich symmetry structures encountered in quantum materials. Combined with the above insights on time crystallization, this naturally leads to $SO(2)\times O(N)$ as the relevant symmetry structure describing non-Abelian time crystals (for $N\geq 3$). 

To disentangle nonlinearities originating from the non-Abelian symmetry structure from those induced by nonequilibrium driving, we systematically construct nonlinear sigma models for the relevant symmetry breaking patterns. In a leading-order loop expansion the renormalization group reveals new fixed points associated with novel universality classes. Remarkably, some of these fixed points exhibit exotic weak scaling behavior, where the $SO(2)$ ``chronon'' associated with broken time translations and the $O(N)$ Goldstone modes display distinct dynamical critical exponents.

\subsection{Synopsis and key results}

\textit{The non-Abelian time crystal group. -- } Our approach is based on only two ingredients: symmetry and the breaking of equilibrium conditions. We therefore expect the resulting phenomenology to arise broadly across very different physical settings. Indeed, $O(N)$ symmetry appears in a wide variety of microscopic models. The enlarged symmetry group $SO(2)\times O(N)$ can emerge along two distinct routes: it may either already be present microscopically, or it may arise effectively at low energies.

The first route naturally occurs in quantum systems with complex symmetry structures. Here, a $U(1)\simeq SO(2)$ factor naturally arises as quantum mechanical phase rotations, while the additional $O(N)$ factor (or related groups such as $SU(2)$) depends on the specific system. In these cases, even weak microscopic driving can activate time-crystalline order; the experimentally observed KPZ phases in exciton-polariton systems~\cite{Fontaine2022,Widmann2026} are examples of this mechanism without the additional $O(N)$ sector. 

The second route arises in systems with only microscopic $O(N)$ symmetry under sufficiently strong driving or pumping. For example, strong parametric drives can induce dynamical order in $O(N)$ models describing, e.g., magnets, thereby generating time-crystalline phases~\cite{Zelle2026,Daviet2024} and naturally leading to the effective field theories studied here. In this case, the $SO(2)$ sector emerges dynamically in the low-frequency rotating-wave approximation.

We also remark that the perhaps more straightforward generalization $SO(2) \to SO(N)$ in lieu of $SO(2) \to SO(2)\times O(N)$ does not lead to interesting new universality classes -- for $N\geq 3$ no non-irrelevant nonequilibrium nonlinearities can be formulated due to the non-Abelian symmetry. It is thus indeed the connection to time-crystallinity that enables the new universality classes discovered here.

\textit{Nonlinear sigma models for rotating and oscillating time crystals. --} The $O(N)\times SO(2)$ field theory admits two distinct ordered phases, corresponding to rotating and oscillating limit cycles. In addition to the Goldstone modes associated with spontaneous $O(N)$ symmetry breaking, both phases contain an $SO(2)$ Goldstone mode describing fluctuations along the limit cycle and thus relate to broken time translations. This ``chronon'' naturally exhibits KPZ nonlinearities \cite{Daviet2025}, but also couples directly to the $O(N)$ Goldstone modes.

Because the broken symmetry is non-Abelian and acts nonlinearly on the Goldstone sector, these interactions must be treated within nonlinear sigma models (NLSMs) defined on the coset space of the broken symmetry group. Equilibrium NLSMs typically have only a single independent coupling leading to leading to a one-parameter  renormalization group flow that determines their phase structure~\cite{ZinnJustin}. Our systematic construction of the nonequilibrium NLSMs for both rotating and oscillating phases reveals that this is different in the $O(N)\times SO(2)$  models out of equilibrium. There are cubic couplings between the chronon and the $O(N)$ Goldstone modes with scaling dimension $\frac{d-2}{2}$, just as the equilibrium coupling. These couplings therefore need to be treated on an equal footing and consequently there is a multiparameter RG flow determining the physics of these NLSMs.

\begin{figure}
    \centering
    \includegraphics[width=1\linewidth]{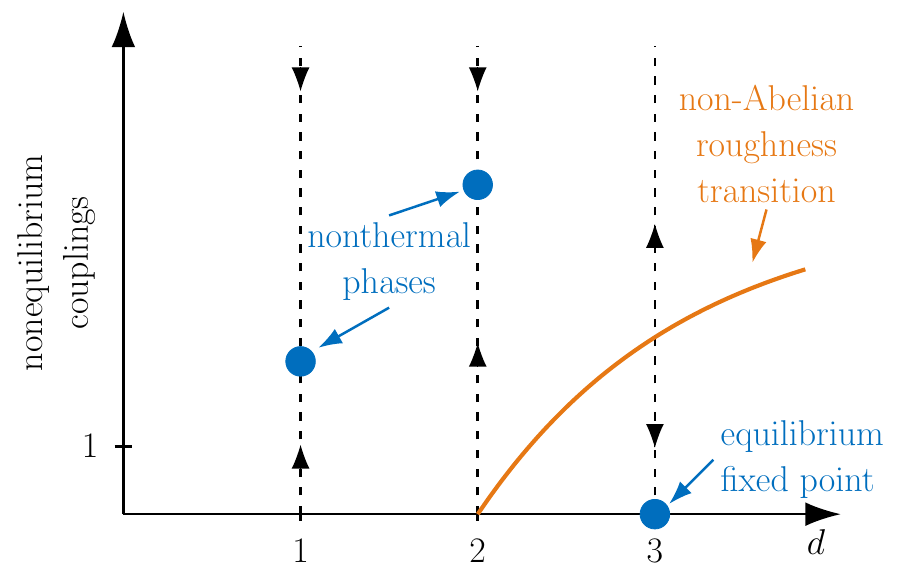}
     \caption{
     Schematic renormalization group flow for the Goldstone modes of the non-Abelian time crystal in various dimensions. In one and two dimensions, the equilibrium axis is unstable: arbitrarily weak nonequilibrium couplings drive the system toward a strongly interacting nonequilibrium fixed point, and no emergent equilibrium exists. In three dimensions, by contrast, the equilibrium line is stable, such that weak nonequilibrium perturbations fade out under coarse graining and effective equilibrium is restored. Above a critical coupling strength, the system undergoes a nonequilibrium phase transition generalizing the three-dimensional KPZ transition to the non-Abelian setting.}
    \label{fig:flow}
\end{figure}

\textit{Universal flow diagram and emergent equilibrium breaking in low dimension. --} To determine the impact of these couplings on the ordered phases of these models, we constrain the model to the deeply ordered limit, where curvature fluctuations are negligible, and expand all couplings to leading order in the Goldstone fields. This reduces the problem to a non-Abelian generalization of KPZ dynamics, which we study in a one-loop RG analysis. In the rotating phase, the model is described by the chronon coupled to $N-2$ complex Goldstone modes, while in the oscillating phase, the coupling is to $N-1$ real Goldstone modes. Beyond establishing these models, a central result is that the flow structure closely parallels the conventional KPZ problem: for arbitrary $N$, the couplings between the chronon and the $O(N)$ modes are relevant in $1+1$ dimensions and marginally relevant in $2+1$ dimensions, see Fig.~\ref{fig:flow}. 

Consequently, all systems governed by $SO(2)\times O(N)$ symmetry exhibit emergent equilibrium breaking: infinitesimal violations of equilibrium conditions become amplified under coarse graining and drive the system toward a strongly coupled nonthermal regime. More operationally in an RG language, the couplings that violate thermal  symmetry~\cite{Sieberer2015,Haehl2016,Glorioso2017,Aron2018} are RG relevant, destabilizing the equilibrium fixed points. The resulting long-distance physics is controlled by genuinely nonequilibrium fixed points without equilibrium counterparts. We determine the crossover scales into this deeply nonthermal regime and characterize the associated scaling behavior analytically and numerically in $1+1$ dimensions. By contrast, in dimensions $d\geq 3$, weak nonequilibrium perturbations become irrelevant and effective equilibrium is restored at long wavelengths.

\textit{Novel scaling behavior: analytical estimates. -- } Despite its nonperturbative nature, we can reach this fixed point in a one-loop renormalization group approach and identify scaling exponents in $1+1$ dimensions. In the rotating phase, the $O(N)$ Goldstones $\vectheta$ describe fluctuations of the rotational orbit. At the fixed point, these modes obey distinct scaling laws from the chronon $\alpha$:
\begin{equation}\label{eq:scaling}
    \mathcal{C}_{\alpha,\theta}(t,\vecr)\sim |\vecr|^{2\chi_{\alpha,\theta}}\hat{\mathcal{C}}_{\alpha,\theta}(|\vecr|^{z_{\alpha,\theta}}/t).
\end{equation}
The critical exponents and universal scaling functions differ between sectors, $ z_\alpha\neq z_\theta$ and $\hat{\mathcal{C}}_{\alpha}(y)\neq\hat{\mathcal{C}}_{\theta}(y)$, while $\chi_\alpha=\chi_\theta$. The most striking signature of this fixed point is therefore the coexistence of distinct dynamical critical exponents within a strongly coupled set of modes. Although the system exhibits universal scaling, the rescaling of time depends on the observable under consideration, a phenomenon known in nonequilibrium statistical mechanics as \textit{weak scaling}~\cite{Taeuber2014}. 

\textit{Novel scaling behavior: Numerical results in $1+1$ dimensions. --  } We complement the analytical analysis by direct numerical simulations of the noisy $O(N)\times SO(2)$ field theory. Since the fixed point lies at strong coupling, quantitative agreement with perturbative RG predictions is not expected. Nevertheless, the simulations provide clear qualitative confirmation of the weak-scaling scenario: the chronon $\alpha$ and the transverse fluctuations $\theta$ indeed display distinct yet universal, parameter-independent scaling functions.

In the oscillating phase, an analogous analysis reduces to the effective theories studied by Erta\c{s} and Kardar~\cite{Ertas1992,Ertas1993} in the context of drifting polymers. Here, the RG flow exhibits four distinct regimes, including KPZ scaling, a second weak-scaling phase, and a nonuniversal unstable regime. We confirm this structure numerically.

\textit{Scaling window and topological defects. --} The above analysis neglects both curvature effects, and the compactness of Goldstone modes, which is associated with the formation of  topological defects that eventually destroy long-range correlations. Similar to the $SO(2)$ exciton-polariton case~\cite{He2015,He2017}, these defects proliferate at the largest scales and ultimately destroy the above scaling, while the new scaling regimes remain observable on experimentally relevant scales in $d=1,2$ \cite{Fontaine2022,Widmann2026}. Previous studies showed that increasing nonequilibrium nonlinearities can trigger a transition 
from a regime with exponentially suppressed vortex densities to one with proliferating defects \cite{He2017}. Our numerical analysis demonstrates that this behavior persists for larger $N$, despite the fact that the relevant topological defects in the $N=3$ rotating phase carry a $\mathbb{Z}_2$ topological charge rather than the $\mathbb{Z}$ charge associated with conventional vortices.

The remainder of this paper is structured as follows. In Sec. \ref{sec:model} we introduce the $O(N)\times SO(2)$ model and systematically derive the effective theories for the Goldstone modes in the respective coset spaces deep in the ordered phase. We then present the one-loop renormalization group analysis of these models in Sec. \ref{sec:rg}. The numerical analysis of the scaling regimes is done in Sec. \ref{sec:num_sim}, while the topological defects are discussed in Sec. \ref{sec:TopDef}.

\section{$O(N)\times SO(2)$ models}\label{sec:model}
\subsection{Model and phase diagram}

Our starting point are $O(N)\times SO(2)$ symmetric field theories for nonequilibrium steady states. $O(N)$ symmetric models arise naturally as the effective field theories of materials with continuous internal symmetries, such as e.g. magnets. The additional $SO(2)$ may exist as a microscopic symmetry. Here, examples are $U(2)\simeq SU(2)\times U(1)$ symmetric systems, such as pumped fermions in electron-hole bilayers, where $U(1)$ and $SU(2)$ describe fermionic phase- and (pseudo-) spin rotations, respectively~\cite{Sun_2024}. 

But it can also be an  emergent symmetry, when $O(N)$ symmetric systems are driven through a driven-dissipative transition, which occurs at a finite frequency~\cite{zelle2024}. In the driven $O(N)$ models, these phases are characterized by limit cycles that either trace out a circular orbit or oscillate along a spontaneously chosen axis. Since these phases break time-translation symmetry $\mathcal{T}\cong\mathbb{R}$ down to discrete time invariance $\mathcal{T}_d\cong\mathbb{Z}$, it breaks the subgroup of time translations  $\mathcal{T}_1\cong\mathbb{R}/\mathbb{Z} \cong SO(2)$. Thus, the long distance physics of a non-Abelian time crystal can be effectively described by a theory that has a $G\times SO(2)$ symmetry~\cite{Cross1993,Daviet2024} where here, we have $G=O(N)$. Since time reversal is also broken in these phases, this symmetry is $SO(2)$ and not $O(2)$. This difference is crucial as it is necessary for KPZ-like physics to emerge \cite{Daviet2025}. It is however also conceivable that the $O(N)\times SO(2)$ symmetry group is implemented directly in the model, i.e. in chiral versions of frustrated and helimagnets that sometimes realize $O(N)\times O(2)$ groups in equilibrium~\cite{Kawamura1998}.

We start by introducing the $O(N)\times SO(2)$ model including nonequilibrium terms. As outlined above and detailed in \cite{Daviet2024} it is the effective field theory of time-crystalline phases of the nonequilibrium $O(N)$ model. We use a complex representation of the order parameter field $\vecpsi \in \mathbb{C}^N$~\cite{Daviet2024}. The $SO(2)$ part of the $O(N)\times SO(2)$ symmetry acts as a $U(1)$ symmetry, while an element $R\in O(N)$ rotates the vector components of $\vecpsi$: 
\begin{align}\label{eq:symmetry_transformation}
    \vecpsi_i \to R_{ij}  \vecpsi_j \exp(i\alpha),  
\end{align}
with $\alpha \in \mathbb{R}$ and $R\in O(N)$. 
The steady state of an open quantum system  out of equilibrium is described by an open-system Keldysh path integral \cite{Sieberer2016, Kamenev2023} that takes the form:
\begin{align}
    Z[\vecj_c,\vecj_q]=\int \mathcal D\vecpsi_c\mathcal D\vecpsi_q  e^{iS[\vecpsi_{q},\vecpsi_c]+\int_{t,\vecr} \vecj_c^\dagger\vecpsi_q+\vecj_q^\dagger\vecpsi_c+c.c.}
\end{align}
Here the classical and quantum fields $\vecpsi_{c,q}\in \mathbb{C}^N$ transform under the symmetry as in \eqref{eq:symmetry_transformation}. For long time scales larger than the inverse noise level one can restrict the field theory to the quadratic level in the quantum fields $\vecpsi_q$ \cite{Sieberer2016,Sieberer2015,Kamenev2023}. This amounts to a semiclassical limit.\\
We split the action into the spectral part, that captures the spectrum of the theory including dissipation and is linear in the quantum fields, and the noise part, that captures the statistical occupation of that spectrum and is quadratic in $\vecpsi_q$ terms,
\begin{align}\label{eq:action}
    S[\vecpsi_q,\vecpsi_c]=S_{\text{spec}}[\vecpsi_q,\vecpsi_c]+S_{\text{noise}}[\vecpsi_q,\vecpsi_c].
\end{align}
For the $O(N)\times SO(2)$ model, the spectral part of the action can be written as
\begin{equation}\label{eq:action_spec}
    S_{\text{spec}}[\vecpsi_q,\vecpsi_c]=\int_{t,\vecr}\vecpsi_q^\dagger\left(\partial_t\vecpsi_c+\frac{\delta H_d}{\delta\vecpsi_c^\dagger}+i\frac{\delta H_c}{\delta\vecpsi_c^\dagger}\right)+c.c.
\end{equation}
with

\begin{align}\label{eq:nonhermitian_hamiltonian}
 H_{c,d} = \int_{\vecr}\boldsymbol{\psi}^*\cdot(-Z_{c,d}\nabla^2 + r_{c,d})\boldsymbol{\psi} + \frac{u_{c,d}}{4} \rho^2 + \frac{\kappa_{c,d}}{4}\tau,
\end{align}
where $\rho$ and $\tau$ are the two invariants of the symmetry group, $\rho=\vecpsi^*\cdot\vecpsi$ and $\tau=\left(\vecpsi\cdot\vecpsi\right)\left(\vecpsi^*\cdot\vecpsi^*\right)$ and $Z_{c,d},r_{c,d},u_{c,d},\kappa_{c,d}\in\mathbb{R}$ are the coupling constants. Note that by virtue of going into a rotating reference frame $\vecpsi_c\rightarrow e^{-ir_ct}\vecpsi_c$ we can always eliminate $r_c$ and henceforth we will set $r_c=0$. This corresponds to an appropriate shift of the parameter $\omega_0$ in \eqref{eq:defphi12}.\\
The noise part of the action generally reads
\begin{align}\label{eq:action_fluc}
    S_{\text{fluc}}=2i\int_{\omega,\vecq}\vecpsi^\dagger_q(\omega,\vecq)\Gamma(\omega,\vecq)\vecpsi_q(\omega,\vecq).
\end{align}

Here, $\int_{\omega,\vecq}=(2\pi)^{-d-1}\int d^dq\,d\omega$. The frequency dependence of $\Gamma(\omega,\vecq)$ captures possible non-Markovian noise contributions. If there is no symmetry or conservation law protecting it, $\Gamma_0=\Gamma(\omega=0,\vecq=0)>0$, so that for the low frequency and wavelength regime that we are interested in, we can set $\Gamma(\omega,\vecq)\approx \Gamma_0$. Note that $\Gamma_0>0$ will be generated by loop corrections, even if it is finetuned to vanish on the level of the bare theory \cite{dalla2012dynamics}. \\
As mentioned above, this model has a phase transition into time-crystalline phases as one tunes $r_d$ through zero. The qualitative  phase diagram as well as the resulting critical theories governing the phase transitions have been worked out in detail in \cite{zelle2024, Daviet2025} and we just briefly review it here. 

Two possible types of long-range order that break $O(N)\times SO(2)$ spontaneously exist, and have $\vecpsi_0(t)=\langle \vecpsi\rangle = \exp(i\omega_0 t)\vecpsi_0\neq 0$ --  Time-translation symmetry is indeed spontaneously broken. It acts on the ground state in the same way as the $SO(2)$ symmetry, as expected from the construction.

 In both phases, $\rho>0$, and these two orders are distinguished by the invariant $\tau$:
\begin{itemize}
    \item  \textbf{Phase A or rotating phase --} For $\kappa_d<0$, $\tau$ vanishes, $\operatorname{Re}{\vecpsi_0}\perp\operatorname{Im}{\vecpsi_0}$ and $||\operatorname{Re}{\vecpsi_0}||=||\operatorname{Im}{\vecpsi_0}||$. 
    \item\textbf{Phase B or oscillating phase -- } For $\kappa_d>0$, $\tau>0$ is maximized, $\operatorname{Re}{\vecpsi_0}||\operatorname{Im}{\vecpsi_0}$.  
    
\end{itemize}

These phases can be connected to the limit cycle phases of the driven $O(N)$ model \cite{zelle2024} via  introducing $\vecpsi=\vecchi_1+i\vecchi_2,\,\vecchi_{1,2}\in\mathbb{R}^N$ and
\begin{equation}
\begin{split}
    \label{eq:defphi12}
    \boldsymbol{\phi}(t,\vecr))&=\boldsymbol{\chi}_1(t,\vecr)\cos(\omega_0t)+\boldsymbol{\chi}_2(t,\vecr)\sin(\omega_0t),\\
    \boldsymbol{\Pi}(t,\vecr)&=-\boldsymbol{\chi}_1(t,\vecr)\sin(\omega_0t)+\boldsymbol{\chi}_2(t,\vecr)\cos(\omega_0t).
    \end{split}
\end{equation}
Here $\vecphi$ is the $O(N)$ order parameter and $\vecpi$ its canonical conjugate. $\omega_0$ is the frequency of the limit cycle and $\vecphi$ is assumed to vary on timescales $\tau^{-1}\ll\omega_0$. The $SO(2)$ symmetry now acts as explicit time-translation. Then, phase A has $\vecphi\perp\partial_t\vecphi$ i.e. corresponds to a rotation on a circle. In phase B $\vecphi\parallel\partial_t\vecphi$, and thus the limit cycle is oscillations along an axis. This motivates the denominations, rotating and oscillating phase. 

\subsection{Effective Field Theories for the Goldstone Modes}

We now construct the effective actions describing the dynamics of the Goldstone modes in phases A and B. The resulting nonlinear sigma models only require knowledge on the symmetry breaking pattern and the nonequilibrium nature of the steady state of the respective phases, and otherwise emerge universally independently of the microscopic details of the problem. 

\subsubsection{Rotating phase}

The ordered phases of the $O(N)\times SO(2)$ models spontaneously break continuous symmetries. Consequently, despite the intrinsically nonequilibrium character of the system, the broken-symmetry phases support Goldstone modes~\cite{zelle2024}. Although these systems do not possess a conserved energy, the Goldstone modes remain soft: their lifetimes diverge algebraically as the wavevector $\vecq$ approaches zero. The underlying mechanism is the nonequilibrium analogue of the familiar equilibrium argument. If a system has a continuous symmetry group $G$ and $\psi_0$ is a steady state, then $g\cdot \psi_0$ is also a steady state for every $g\in G$. When the steady state spontaneously breaks the symmetry, namely when $g\cdot \psi_0\neq \psi_0$ for some $g\in G$, the system admits a continuous manifold of nonequilibrium steady states. Denoting by $H$ the unbroken subgroup that leaves the steady state invariant, this manifold is the coset space $G/H$, and fluctuations tangent to this manifold constitute the soft Goldstone modes. We now systematically construct the effective nonequilibrium field theory, or nonlinear sigma model, associated with the relevant cosets of the $O(N)\times SO(2)$ model.

We begin by identifying the continuous symmetries that are broken in the rotating phase. Without loss of generality, we choose the order parameter to be
\begin{align}
    \label{eq:rot_ground_state}
    \vecpsi_{0,A}= e^{i\omega_0 t}\sqrt{\frac{\rho_0}{2}}\left(\hat{e}_1+i\hat{e}_2\right).
\end{align}
This phase breaks the symmetry group $G=O(N)\times SO(2)$ down to the unbroken subgroup
\begin{equation}
    H=O(N-2)\times SO(2)_d .
\end{equation}
The $O(N-2)$ factor rotates the $N-2$ vanishing components of $\vecpsi_{0,A}$ among themselves. The diagonal subgroup $SO(2)_d$ consists of a complex phase rotation by an angle $\alpha$ combined with an $O(N)$ rotation in the $(1,2)$ plane by an angle $-\alpha$.

The coset therefore contains $2N-3$ Goldstone modes, associated with the broken generators $T^{(1,i)}$ and $T^{(2,i)}$ for $i=3,\ldots,N$, together with
\begin{equation}
    T^\alpha \equiv T^{(1,2)}-i\mathbb{1}_2 ,
\end{equation}
where
\begin{equation}
    \left[T^{(i,j)}\right]_{ab} = \delta^j_a\delta^i_b-\delta^i_a\delta^j_b .
\end{equation}
This symmetry breaking pattern has a distinctive structural feature: among all broken generators, the single generator $T^{(1,2)}-i\mathbb{1}_2$ is distinguished by commuting with the unbroken subgroup $H$. It generates fluctuations along the limit cycle, and its associated Goldstone field $\alpha$ transforms as a scalar in the effective field theory. This mode is the chronon, the Goldstone mode associated with broken time translations. As a result, the effective theory admits KPZ-type nonlinearities that drive the dynamics genuinely far from equilibrium. Such interactions are typically forbidden in equilibrium symmetry breaking patterns, such as those of the $O(N)/O(N-1)$ nonlinear sigma model.

The remaining Goldstone modes may be organized as fields $\theta_i^a$, with $a=1,2$ and $i=3,\ldots,N$, corresponding to the broken generators $T^{(a,i)}$. These fields transform covariantly under the unbroken subgroup according to
\begin{equation}
    \theta_i^a \rightarrow S_{ab}R^{ij}\theta_j^b,\quad
    S\in SO(2)_d,\quad R\in O(N-2).
\end{equation}
They describe angular fluctuations of the limit-cycle orbit itself.

\begin{figure}
    \begin{minipage}[b!]{0.6\columnwidth}
        \centering
        \includegraphics[width=\columnwidth]{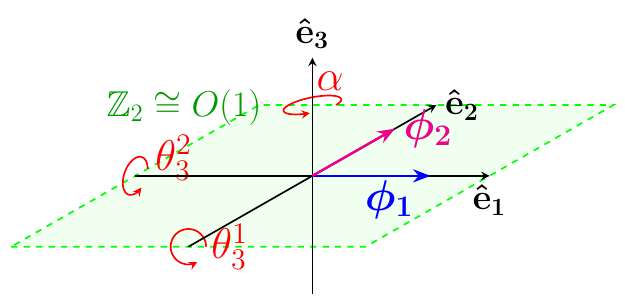}
        \text{(a)}
    \end{minipage}
    \begin{minipage}[b!]{0.38\columnwidth}
    \centering
        \includegraphics[width=\columnwidth]{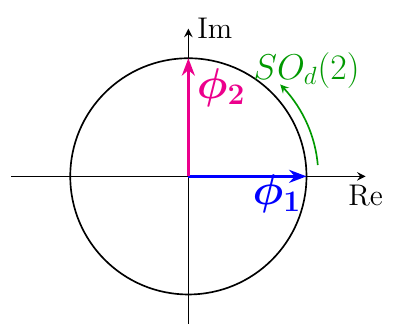}
        \text{(b)}
    \end{minipage}

    \caption{Rotating phase Goldstone modes for $N=3$. (a) depicts how the Goldstone modes in $O(3)/O(1)$ act on the ground state $\vecpsi_0=e^{i\omega_0 t}(\boldsymbol{\phi}_1+i\boldsymbol{\phi}_2)$. The unbroken $O(1)$ symmetry acts as reflection in the $\mathbf{\hat{e}_1}$-$\mathbf{\hat{e}_2}$ plane. A rotation by $\alpha$ in combination with an opposite rotation in the real-imaginary plane (b) leave the ground state invariant and hence form the unbroken symmetry group $SO_d(2)$. Additionally, the time dependence $e^{i\omega_0 t}$ rotates $\mathrm{Re}(\vecpsi_0)$ around the $\mathbf{\hat{e}_1}$-$\mathbf{\hat{e}_2}$ plane and hence the phase is referred to as rotating.}
    \label{fig:rotating_goldstones}
\end{figure}

The effective action on these coset spaces can be derived using the coset construction, originally developed in ref.~\cite{Callan1969} and extended to nonequilibrium effective actions following the formalism of ref.~\cite{Hongo2021}. The construction relies only on two requirements: the full action must be invariant under the microscopic symmetry group $G$, and the Goldstone fields must transform linearly under the unbroken subgroup $H$.

To describe fluctuations within the rotating phase, we therefore seek an effective Keldysh action for the modes parameterizing the coset
\begin{align}\label{eq:coset}
    \frac{O(N)\times SO(2)}{O(N-2)\times SO(2)_d}
    \cong
    \frac{O(N)}{O(N-2)} .
\end{align}
The latter simplification~\cite{Azaria1993} allows us to work with purely real representations rather than mixing real rotations and complex phases as above.
We use the steady state in Eq.~\eqref{eq:rot_ground_state} as a reference point and parametrize the coset by acting on it with symmetry transformations generated by the broken directions:
\begin{equation}
    \label{eq:coset_parametrization}
    \begin{split}
        \vecpsi_{G/H}(t,\vecr)&=U(\vecPhi(t,\vecr))\vecpsi_{0,A} =\exp\!\left[\vecPhi(t,\vecr)\right]\vecpsi_{0,A},
        \\
        \vecPhi(t,\vecr)&=\alpha(t,\vecr)T^{(1,2)}+\sum_{a=1,2}\sum_{i=3}^{N}\theta_i^a(t,\vecr)T^{(a,i)}
        \\&\in\mathcal{L}\!\left[O(N)\right]-\mathcal{L}\!\left[O(N-2)\right] .
    \end{split}
\end{equation}
Here, $\mathcal{L}(G),\mathcal{L}(H)$ are the Lie algebras of the groups $G$ and $H$ respectively and $\alpha(t,\vecr)$ and $\theta_i^a(t,\vecr)$ are the Goldstone fields associated with the broken generators. They take values in the tangent space of the target manifold $G/H$, namely in the complement of $\mathcal{L}(H)$ inside $\mathcal{L}(G)$. In the present case, $G=O(N)$ and $H=O(N-2)$ after quotienting by the diagonal $SO(2)_d$ redundancy. Equation~\eqref{eq:coset_parametrization} defines a local parameterization of the coset representative $U\in G/H$. This parameterization depends on the choice of reference steady state in Eq.~\eqref{eq:rot_ground_state}. Moreover, it contains the usual coset redundancy: $U(\vecPhi)$ and $U(\vecPhi)h$ represent the same physical point on $G/H$ for any $h\in H$.

The fundamental building block for the derivative expansion is the Maurer--Cartan form
\begin{align}
    \vecbeta^\mu:=
    U(\vecPhi)^{-1}\partial^\mu U(\vecPhi)\in\mathcal{L}(G)
\end{align}
where $\partial_\mu$ collects the time and space derivatives in $d+1$ dimensional Euclidean spacetime. Although $\vecPhi$ itself contains only broken generators, $\vecbeta^\mu$ generally has components along both broken and unbroken directions. This is because $\vecPhi(t,\vecr)$ and $\partial_\mu\vecPhi(t,\vecr)$ need not commute. Equivalently, as one moves on the curved coset manifold, the local tangent frame rotates, and this rotation can have components in the unbroken algebra.

We therefore decompose the Maurer--Cartan form into a component tangent to the coset and a component in the unbroken algebra,
\begin{align}
    \vecbeta^\mu=\vecbeta_\perp^\mu +\vecbeta_\parallel^\mu,
    \quad\vecbeta_\perp^\mu\in\mathcal{L}(G)-\mathcal{L}(H),
    \quad\vecbeta_\parallel^\mu\in \mathcal{L}(H).
\end{align}
The transverse component $\vecbeta_\perp^\mu$ contains the physical soft Goldstone gradients. It is useful to further separate it according to the distinguished chronon direction and the remaining angular fluctuations of the limit-cycle orbit:
\begin{align}
    \vecbeta_\perp^\mu=\vecbeta_\alpha^\mu+\vecbeta_\theta^\mu .
\end{align}
The component $\vecbeta_\alpha^\mu$ is associated with motion along the limit cycle and is invariant under the unbroken subgroup, whereas $\vecbeta_\theta^\mu$ transforms covariantly under $H$. We write
\begin{align}
    \label{eq:vielbein}
    \vecbeta_\alpha^\mu&=f^{\alpha,\mu}(\alpha,\theta_i^a)T^{(1,2)},
    \\
    \vecbeta_\theta^\mu&=\sum_{a=1,2}\sum_{i=3}^{N}f_{a,i}^{\theta,\mu}(\alpha,\theta_i^a)T^{(a,i)} .
\end{align}
The functions $f^{\alpha,\mu}$ and $f_{a,i}^{\theta,\mu}$ are the vielbeins on the coset manifold. They encode the nonlinear relation between the coordinates $(\alpha,\theta_i^a)$ and the physical tangent vectors on $G/H$. Their field dependence is completely fixed by the symmetry breaking pattern.

This geometric structure also explains why ordinary derivatives are not sufficient beyond leading order. Since $\vecbeta_\perp^\mu$ is a tangent vector on a curved target manifold, differentiating it at two nearby spacetime points requires comparing tangent vectors defined in different local frames. The unbroken component $\vecbeta_\parallel^\mu$ precisely supplies the compensating local $H$ rotation between these frames. It therefore plays the role of the connection associated with the coset geometry. The covariant derivative acting on the Goldstone gradients is consequently
\begin{align}
    \nabla^\mu \vecbeta_\perp^\nu=\partial^\mu \vecbeta_\perp^\nu+\left[\vecbeta_\parallel^\mu,\vecbeta_\perp^\nu\right].
\end{align}
This derivative transforms covariantly under the unbroken subgroup and is therefore the appropriate building block for constructing higher-gradient terms in the effective action. Importantly, both $\vecbeta^\nu_\theta$ as well as $\nabla^\nu\vecbeta_\theta^\mu$ transform in a vector representation of $H$ while the chronon $\vecbeta_\alpha^\mu$ and its covariant derivative are singlets in $H$.

So far, the construction parallels the standard coset construction for the target space $O(N)/O(N-2)$. In a statistical mechanics nonlinear sigma model, the leading invariant Lagrangian would simply be
\begin{equation}
    L_{\mathrm{eq}}=\operatorname{Tr}\left[\vecbeta_\perp^\mu \vecbeta_{\perp,\mu}\right],
\end{equation}
up to model-dependent stiffnesses. This Lagrangian captures the stationary state of equilibrium dynamics. Here we are interested in the more general nonequilibrium situation, where an \emph{a priori} reduction to a static description is not possible in general. Instead we need the dynamical Keldysh formulation able to capture nonequilibrium situations. In particular, this formulation contains the corresponding quantum, or response, fields. The latter encode the response of the stochastic dynamics to perturbations and generate fluctuations along the same coset directions as the Goldstone modes \cite{Hongo2021}. Since the physical fluctuations are tangent to the coset manifold at the ordered state, it is natural to represent the response fields as algebra-valued tangent vectors. We therefore decompose them into the chronon response and the transverse angular responses,
\begin{align}
    \vecpsi_q&=U(\tilde{\boldsymbol{\alpha}}+\tilde{\boldsymbol{\Theta}})\vecpsi_0,
    \\\tilde{\boldsymbol{\alpha}}&=\tilde\alpha\, T^{(1,2)},
    \\
    \tilde{\boldsymbol{\Theta}}&=\sum_{a=1,2}\sum_{i=3}^{N}
    \tilde\theta_{a,i}\,T^{(a,i)} .
\end{align}
The response fields transform like their their counterparts: The chronon response is an $H$ singlet, while the transverse response fields transform covariantly under the group $H$:
\begin{equation}
    \begin{aligned}
        \tilde{\boldsymbol{\alpha}}
        &\rightarrow\tilde{\boldsymbol{\alpha}},\\
        \tilde{\boldsymbol{\Theta}}&\rightarrow h\,\tilde{\boldsymbol{\Theta}}\,h^{-1}.
    \end{aligned}
\end{equation}
This makes the response fields the Keldysh counterparts of the covariant Goldstone gradients $\vecbeta_\alpha^\mu$ and $\vecbeta_\theta^\mu$. Since the response fields enter the leading Keldysh action without derivatives, no additional covariant derivative acting on response fields is required at this order.

We may now construct the most general nonrelativistic Keldysh Lagrangian for the coset
\begin{equation}
    \frac{O(N)\times SO(2)}{O(N-2)\times SO(2)_d}
\end{equation}
containing at most one time derivative and two spatial derivatives by collecting all invariant contractions of the Goldstone fields and their responses. The result is 
\begin{widetext}
\begin{equation}\label{eq:nlsm_rot}
    \begin{aligned}
        {L_A}=\operatorname{Tr} \Big[&P\tilde{\boldsymbol{\Theta}}^T\left(\vecbeta_\theta^0-Z_\theta\nabla^i\vecbeta_\theta^i+ig\vecbeta_\perp^i\vecbeta_\perp^i\right)+\mathrm{c.c.}+2i\Gamma_\theta P\tilde{\boldsymbol{\Theta}}^T\tilde{\boldsymbol{\Theta}}
        \\&2P\tilde{\boldsymbol{\alpha}}^T\left(\vecbeta_\alpha^0-Z_\alpha\nabla^i\vecbeta_\alpha^i-i\lambda_\alpha\vecbeta^i_\alpha\vecbeta^i_\alpha-i\lambda_\theta\vecbeta_\theta^i\vecbeta_\theta^i\right)+2i\Gamma_\alpha P\tilde{\boldsymbol{\alpha}}^T\tilde{\boldsymbol{\alpha}}\Big],
    \end{aligned}
\end{equation}

\end{widetext} 
where the projector, $P=\frac{1}{2}(\mathbb{1}-iT^{(1,2)})$ onto the ground state $\vecpsi_0$ arises from the scalar product being written as a trace and allows us to only consider covariant contributions to the action. The coefficients $Z_\theta=Z_d+iZ_c$ and $g=g'+ig''$ are complex and all other coefficients are real.

Equation~\eqref{eq:nlsm_rot} is the nonequilibrium nonlinear sigma model for the Goldstone modes in the rotating phase A. Its structure reflects the distinction between the scalar chronon sector, and the transverse sector transforming under the unbroken subgroup. In particular, the nonlinear terms proportional to $\lambda_\alpha$, $\lambda_\theta$, and $g', g''$ are allowed by symmetry in the nonequilibrium theory and encode the leading KPZ-type couplings between gradients of the Goldstone fields. All of them have scaling dimension $\frac{d-2}{2}$, like the equilibrium sigma model coupling. This nonequilibrium NLSM therefore has a multicomponent RG flow, in stark contrast the the single parameter RG of equilibrium NLSMs~\cite{ZinnJustin}.

We now focus on the regime deep inside the ordered phase, far from the critical point. In this limit, the target-space curvature produces only higher-order interaction vertices, and the leading infrared dynamics is obtained by expanding the Maurer--Cartan forms to first order in the Goldstone fields. Equivalently, one approximates the coset manifold locally by its tangent plane at the ordered state. This gives
\begin{align}
    \vecbeta_\alpha^\mu &\simeq\partial^\mu\alpha\,T^{(1,2)},
    \\
    \vecbeta_\theta^\mu&\simeq\sum_{a=1,2}\sum_{i=3}^{N} \partial^\mu\theta_i^a\,T^{(a,i)}.
\end{align}
We keep the leading order contributions of all the independent terms in the original action Eq.~\eqref{eq:nlsm_rot}.
To streamline the notation, we combine the transverse Goldstone fields into complex vectors,
\begin{equation}
    \begin{aligned}
        \vectheta&=\frac{1}{\sqrt{2}}\left(\theta_3^1+i\theta_3^2,\ldots,\theta_N^1+i\theta_N^2\right)\in\mathbb{C}^{N-2},
        \\
        \tilde{\vectheta}&=\frac{1}{\sqrt{2}}\left(\tilde\theta_3^1+i\tilde\theta_3^2,\ldots,\tilde\theta_N^1+i\tilde\theta_N^2\right)\in\mathbb{C}^{N-2}.
    \end{aligned}
\end{equation}
Inserting this leading-order vielbein expansion into Eq.~\eqref{eq:nlsm_rot} yields the flat-target-space Goldstone action
\begin{equation}
    \label{eq:rot_goldstone_action}
    \begin{split}
        S_A^{\mathrm{flat}}=\int_{t,\vecr}&2\tilde\alpha\left[\partial_t\alpha-Z_\alpha\nabla^2\alpha+\lambda_\alpha(\nabla\alpha)^2+\lambda_\theta|\nabla\vectheta|^2\right]
        \\
        &+\tilde{\vectheta}^{\,*}\cdot\left[\partial_t\vectheta-Z_\theta\nabla^2\vectheta+g\,(\nabla\alpha)\cdot(\nabla\vectheta)\right]+\mathrm{c.c.}
        \\
        &+2i\Gamma_\alpha\tilde\alpha^2+2i\Gamma_\theta|\tilde{\vectheta}|^2 .
    \end{split}
\end{equation}

This action describes the local limit of the full nonlinear sigma model where the target space's curvature is neglected. In this approximation, all interaction vertices originating solely from the curvature of the Goldstone manifold are neglected, while the symmetry-allowed nonequilibrium gradient nonlinearities are retained. The couplings in Eq.~\eqref{eq:rot_goldstone_action} can be related to the parameters of the microscopic action in Eq.~\eqref{eq:nonhermitian_hamiltonian} by expanding in phase and amplitude fluctuations around the mean-field solution in Eq.~\eqref{eq:rot_ground_state}; the details are given in Appendix~\ref{app:phase_amplitude}.

This contains the first key result of this paper. The cubic terms of this Lagrangian can only exist because of the existence of the $SO(2)$ Goldstone mode of time translation, the chronon $\alpha,\tilde\alpha$, that is a scalar under the unbroken subgroup and further carries no conjugation symmetry - there is no symmetry between the fluctuations along and against the limit cycles direction. The coupling proportional to $\lambda_\alpha$ indeed reduces to the KPZ-coupling of the pure $U(1)$ Goldstone mode in the flat limit, which destabilizes the Gaussian theory and pushes it to a strongly coupled, nonthermal fixed point. This is confirmed in theory \cite{Altman2015, Daviet2025}, numerics \cite{Helluin2025, Deligiannis2022} and experiment \cite{Fontaine2022,Widmann2026}. The couplings proportional to $\lambda_\theta$ and $g$ now provide the generalizations of this to larger, non-Abelian symmetry groups. We stress that this result arises solely due to symmetry, and by allowing nonequilibrium couplings. As we will see below, already a leading order expansion of this model leads to novel scaling structures in low dimensions.\\

 These couplings are firmly tied to the nonequilibrium nature of the problem. An (effective) equilibrium can be associated with the presence of a thermal time reversal  symmetry of the Keldysh action that ensures fluctuation-dissipation relations (FDR) of the corresponding responses and correlations.  This is ensured, if the corresponding Keldysh action has a thermal time-reversal symmetry \cite{Sieberer2015,Haehl2016,Glorioso2017,Aron2018}. The original action Eq.~\eqref{eq:action} is at thermodynamic equilibrium for $H_c= kH_d$, such that $k=\frac{\mathrm{Im}(Z_\theta)}{\mathrm{Re}(Z_\theta)}$. This implies that the thermal time reversal symmetry for the NLSM action Eq.(\ref{eq:nlsm_rot}) takes the form 
\begin{equation}
    \begin{aligned}
    \vecbeta^\mu(t,\vecr)&\to\vecbeta^{\mu}(-t,\vecr)^T=-\vecbeta^{\mu}(-t,\vecr),\\ 
        \tilde{\boldsymbol{\Theta}}(t,\vecr)P&\to\frac{1}{1-ik}P\left((1+ik)\tilde{\boldsymbol{\Theta}}^T(-t,\vecr)-\frac{i}{\Gamma_\theta}\beta_\theta^{0}(-t,\vecr)\right),\\
       P\tilde{\boldsymbol{\Theta}}^T(t,\vecr)&\to\frac{1}{1+ik}\left((1-ik)\tilde{\boldsymbol{\Theta}}(-t,\vecr)+\frac{i}{\Gamma_\theta}\beta_\theta^{0}(-t,\vecr)\right)P,\\
        \tilde{\boldsymbol{\alpha}}(t,\vecr)&\to \tilde{\boldsymbol{\alpha}}^T(-t,\vecr)-\frac{i}{\Gamma_\alpha}\vecbeta_\alpha^0(-t,\vecr).
    \end{aligned}    
\end{equation} 

From this full transformation, we conclude that at equilibrium $\lambda_\theta=\lambda_\alpha=0$, $\frac{\mathrm{Re}(Z_\theta)}{\Gamma_\theta}=\frac{Z_\alpha}{\Gamma_\alpha}$ and $g=iZ_\theta$. 

In the flat limit Eq.~\eqref{eq:rot_goldstone_action} the symmetry transformation is linearly realized and reads

\begin{equation}\label{eq:thermal_symmetry_flat}
    \begin{aligned}
        \boldsymbol{\theta}(t,\vecr)&\to\boldsymbol{\theta^*}(-t,\vecr),\\
        \boldsymbol{\theta^*}(t,\vecr)&\to\boldsymbol{\theta}(-t,\vecr),\\
        \alpha(t,\vecr)&\to- \alpha(-t,\vecr),\\
        \tilde{\boldsymbol{\theta}}(t,\vecr)&\to\frac{1}{1-ir}\Big[(1+ir)\tilde{\boldsymbol{\theta^*}}(-t,\vecr)+\frac{i}{\Gamma_\theta}\partial_t\boldsymbol{\theta^*}(-t,\vecr)\Big],\\
        \tilde{\boldsymbol{\theta^*}}(t,\vecr)&\to\frac{1}{1+ir}\Big[(1-ir)\tilde{\boldsymbol{\theta}}(-t,\vecr)+\frac{i}{\Gamma_\theta}\partial_t\boldsymbol{\theta}(-t,\vecr)\Big] \\
        \tilde{\alpha}(t,\vecr)&\to-\tilde{\alpha}(-t,\vecr)-\frac{i}{\Gamma_\alpha}\partial_t\alpha(-t,\vecr).
    \end{aligned}
\end{equation}
We observe that the non-linearities $\lambda_\alpha, \lambda_\theta$ and $g$  break this symmetry explicitly in the flat limit and are thus identified as non-thermal. 

There is a subtlety in the one dimensional case, which already appears in the ordinary KPZ equation~\cite{Kamenev2023}. While the transformation above is not generally a symmetry of the problem, and the dynamical actions above break equilibrium in this sense, there is a Gaussian stationary probability distribution solving the associated Fokker-Planck equation, just as for the stochastic diffusion problem at equilibrium (see~\cite{Frey1996,Canet2011} for a modified capturing this behavior). We find such a Gaussian equilibrium distribution for $\mathrm{Im}(g)=0$ and $\frac{\lambda_\theta Z_\alpha}{\Gamma_\alpha}=\frac{\mathrm{Re}(g)\mathrm{Re}(Z_\theta)}{\Gamma_\theta}$ for any value of $\lambda_\alpha$ (see appendix \ref{app:fokker_planck}). 
In the following section, we show that $\mathrm{Im}(g)\neq0$ at the fixed point governing this dynamical system. This evidences a genuine far-from-equilibrium system with a genuinely non-thermal distribution in one dimension, in stark contrast to the ordinary KPZ behavior, as per the above discussion.

\subsubsection{Oscillating phase}
We now repeat this construction for the oscillating phase or phase B, where we write the steady state without loss of generality as
\begin{align}
    \vecpsi_{0,N}=e^{i\omega_0 t}\sqrt{\rho_0}\hat{e}_1.
\end{align}
This order breaks the $ SO(2)$ symmetry fully, and $O(N)$ to $O(N-1)$. The complex phase and the angular fluctuations in $O(N)/O(N-1)$ cleanly separate on the algebraic level. We then have an $SO(2)$ chronon coupling to the modes on the $O(N)/O(N-1)$ coset, which is the familiar one from symmetry breaking in the $O(N)$ model.\\

\begin{figure}
    \begin{minipage}[b!]{0.5\columnwidth}
        \centering
        \includegraphics[width=\columnwidth]{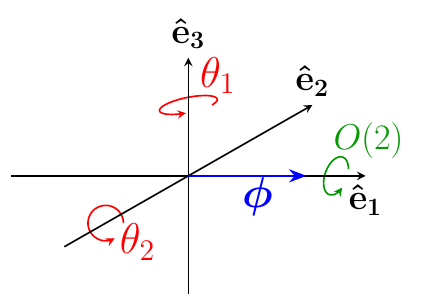}
        \text{(a)}
    \end{minipage}
    \begin{minipage}[b!]{0.48\columnwidth}
    \centering
        \includegraphics[width=\columnwidth]{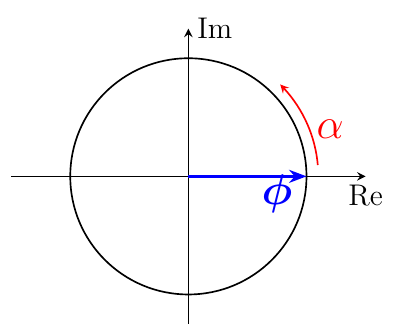}
        \text{(b)}
    \end{minipage}

    \caption{
    Oscillating phase Goldstone modes for $N=3$. (a) illustrates how the Goldstone modes in $SO(2)\times O(3)/O(2)$ act on the ground state $\vecpsi_0=e^{i\omega t}\boldsymbol{\phi}$. The unbroken $O(2)$ symmetry acts as an $SO(2)$ rotation around $\mathbf{\hat{e}_1}$ and a $\mathbb{Z}_2$ reflection in the $\mathbf{\hat{e}_2}$-$\mathbf{\hat{e}_3}$ plane. The Goldstone mode $\alpha$ (b) parametrizes the broken $SO(2)$ symmetry equivalent to rotation in the real-imaginary plane. A rotation by $\alpha$ in combination with an opposite rotation in the real-imaginary plan (b) leave the ground state invariant and hence form the unbroken symmetry group $SO_d(2)$. Additionally, the time dependence makes $\mathrm{Re}(\vecpsi_0)$ oscillate in magnitude along the $\mathbf{\hat{e}_1}$ direction and hence is named the oscillating phase.}
    \label{fig:oscilating_goldstones}
\end{figure}

The Maurer-Cartan form for the fluctuations on the $O(N)/O(N-1)$ coset has the components
\begin{align}
    \vecbeta_\theta^\mu=\sum_{i=2,...,N}f^\mu_a(\vectheta)T^{(1,i)}
\end{align}
with $f^\mu_i(\vectheta)\approx\partial_\mu\theta_i$, and the corresponding response field reads
\begin{align}
    \tilde{\boldsymbol{\Theta}}=\sum_{i=2,...,N}\tilde\theta_iT^{(1,i)}.
\end{align}
This then allows us to construct the invariant Keldysh Lagrangian for the Goldstone modes in phase B
\begin{equation}\label{eq:nlsm_osc}
    \begin{aligned}
    {L_A}=\operatorname{Tr} \Big[
        &\tilde\alpha\left(\partial_t\alpha-Z_\alpha\nabla^i\nabla^i\alpha+\lambda_\alpha(\nabla^i\alpha)^2-\frac{\lambda_\theta}{2}\vecbeta_\theta^i\vecbeta_\theta^i\right)\\
        &\tilde{\boldsymbol{\Theta}}^T\left(\vecbeta_\theta^0-Z_\theta\nabla^i\vecbeta_\theta^i+g\vecbeta_\theta^i\nabla^i\alpha\right)\\&+i\Gamma_\alpha\tilde\alpha\tilde\alpha+i\Gamma_\theta\tilde{\boldsymbol{\Theta}}^T\tilde{\boldsymbol{\Theta}}\Big].
    \end{aligned}
\end{equation}
By neglecting the angular fluctuations $\theta$, we recover the KPZ behavior of the chronon $\alpha$. On the other hand without the chronon mode, the Lagrangian becomes Gaussian and resembles the diffusive Goldstone modes of an equilibrium $O(N)$ model.

Again expanding to leading order gives  
\begin{equation}\label{eq:osc_goldstone_action}
\begin{split}
    S_{B}^{\text{flat}}=\int_{t,\vecr}&\tilde\alpha\left(\partial_t\alpha-Z_\alpha\nabla^2\alpha+\lambda_\alpha(\nabla\alpha)^2+\lambda_\theta|\nabla\vectheta|^2\right)\\
    &+\tilde\vectheta^T\cdot\left(\partial_t\vectheta-Z_\theta\nabla^2\vectheta+g\nabla\alpha\nabla\vectheta\right)\\
    &+i\Gamma_\alpha\tilde\alpha^2+i\Gamma_\theta|\tilde\vectheta|^2,
    \end{split}
\end{equation}
where $\tilde\vectheta,\vectheta\in\mathbb{R}^{N-1}$ and $Z_\theta,g$ are now real valued. This action also contains a KPZ nonlinearity as expected and two rather than three additional couplings to the angular fields. In this flat limit, our model corresponds to the Langevin dynamics studied in \cite{Ertas1992,Ertas1993}.

By imposing the thermal symmetry via the following transformation:
\begin{equation}
    \begin{aligned}
    \alpha(t,\vecr)&\to-\alpha(-t,\vecr)\\
    \vecbeta_{\theta}^\mu(t,\vecr)&\to\vecbeta_{\theta}^\mu(-t,\vecr)^T=-\vecbeta_{\theta}^\mu(-t,\vecr)\\ 
        \tilde{\boldsymbol{\Theta}}(t,\vecr)&\to-\tilde{\boldsymbol{\Theta}}(-t,\vecr)-\frac{i}{\Gamma_\theta}\beta_\theta^{0}(-t,\vecr)\\
        \tilde{{\alpha}}(t,\vecr)&\to -\tilde{{\alpha}}(-t,\vecr)-\frac{i}{\Gamma_\alpha}\partial_t\alpha(-t,\vecr),
    \end{aligned}    
\end{equation}
conclude that at equilibrium, $g=\lambda_\alpha=\lambda_\theta=0$.

In $d=1$ dimension there is a Gaussian  equilibrium stationary distribution for $2\lambda_\theta Z_\alpha\Gamma_\theta=gZ_\theta\Gamma_\alpha$ as shown in \cite{Ertas1993}.

\section{Scaling and renormalization group analysis} \label{sec:rg}

Since the Goldstone modes are gapless by nature, there is an infinite correlation length and their long wavelength behavior will be captured by a renormalization group fixed point of the respective nonlinear sigma models \eqref{eq:nlsm_rot} and \eqref{eq:nlsm_osc}. We do a perturbative analysis around the Gaussian fixed point to determine when the nonthermal couplings push the system to an interacting fixed point. 

To that end, we make the following scaling ansatz for the Goldstone mode correlators in analogy to the original KPZ problem
\begin{align}
\label{eq:2pointCorr}
    \langle(\alpha(t,\vecr)-\alpha(0,0))^2\rangle&=\mathcal{C}_{\alpha}(t,\vecr)\sim |\vecr|^{2\chi_\alpha}\hat{\mathcal{C}}_\alpha(|\vecr|^{z_\alpha}/t)\\
    \langle|\theta_i(t,\vecr)-\theta_j(0,0)|^2\rangle&=\mathcal{C}_{\theta}(t,\vecr)\delta_{ij}\sim |\vecr|^{2\chi_\theta}\hat{\mathcal{C}}_\theta(|\vecr|^{z_\theta}/t)\delta_{ij},
\end{align}
where $i,j=1,...,N-2$.
At the non-interacting level one has Edward-Wilkinson (EW) scaling
\begin{align}
    z=2,\,\chi= 1-\frac{d}{2}.
\end{align}
We already anticipate that the corrections due to the interactions will destroy the EW scaling and  lead to \emph{weak scaling}, where the scaling exponents for the two types of Goldstone modes differ. In particular, there is no unique dynamical critical exponent $z$.\\

In the following, we are interested in the physics deep within the ordered phase, far from the critical point. We therefore can restrict ourselves to the flat approximations \eqref{eq:rot_goldstone_action} and \eqref{eq:osc_goldstone_action} respectively, which will give us information about scaling behavior associated with smooth nonequilibrium fluctuations. This discards the compact nature of the angular fields $\alpha,\theta$, which asymptotically leads to the formation of topological defects, giving rise to a cutoff for the above scaling behaviors; we discuss these in Sec. \ref{sec:TopDef}. To derive the renormalization group flow equations, we need to calculate the corrections to all the couplings occurring in the action. The contributions from $\lambda_\alpha,\lambda_\theta$ and $g$ all stem from vertex corrections. The corrections to the stiffness parameters $Z_\alpha$ and $Z_\theta$ originate from the retarded self energy $\lim_{\vecq\rightarrow 0}\partial_{\vecq^2}\Sigma^R(\omega=0,\vecq)$, and the noise level corrections to $\Gamma_{\alpha,\theta}$ from the Keldysh self energies $\Sigma^K(\omega=0,\vecq=0)$ in presence of a momentum shell cutoff at a scale $\Lambda$. The first important observation is that the vertex corrections scale as $\Lambda^{d-2}$. Therefore, they are relevant perturbations in $d<2$, and turn out to be marginally relevant in $d=2$. This entails that for $d\leq2$ the system always flows to an interacting fixed point while there is a transition between a free, diffusive regime to a roughly fluctuating one at a finite nonlinearity strength for $d>2$ (see Fig.~\ref{fig:flow}). This qualitatively resembles the phase diagram of the pure KPZ equation. The fixed points themselves, as well as the scaling associated to them, are drastically different as we will see below. 
\\
 We compute the dimensionful renormalization group $\beta$-functions in a Wilsonian momentum shell scheme, taking a derivative of the self-energy and vertex corrections with respect to the UV cutoff $\Lambda$. The details of this derivation including all diagrams and the full set of dimensionful flow equations are given in appendix \ref{app:loops}. The most important qualitative aspect, which underlies the novel scaling behavior established below, can be gleaned from the dimensionless flow equation of the KPZ coupling $\lambda_\alpha$: 
\begin{align}\label{eq:KPZ_rotating_dimless}
\Lambda\partial_\Lambda\lambda_\alpha=&\Lambda^{d-2}\frac{4(N-2)\Gamma_\theta {g''}^2\lambda_\theta}{Z_d^3}.
\end{align}
Note that in the case of pure KPZ ($g''=\lambda_\theta=0$), the dimensionful $\beta$ function of this coupling vanishes. This is a consequence of the Galilean shift symmetry of the KPZ equation \cite{Kloss2012};  it is the dimensionless KPZ coupling  involving noise level and diffusion constant that flows and gives rise to the KPZ flow phenomenology. The vanishing remains true in the oscillating phase, where
the coupling $g'' =0$ and the system can flow to a decoupled KPZ fixed point \cite{Ertas1992}. We can however see immediately, that any finite $g''$ will change this behavior qualitatively and push the system to a novel fixed point. We note that the rotating phase only has this nontrivial structure if $N\geq 3$, i.e. in the case of a non-Abelian continuous symmetry. For $N=1$ there is no rotating phase and for $N=2$ there is only one Goldstone mode which will show KPZ behavior.\\

\subsection{Flow equations and scaling in the rotating phase}

We now proceed to the study of the scaling fixed point of the RG flows in the rotating phase. To that end, we  define the following dimensionless couplings at a renormalization scale $\Lambda$
\begin{equation}
\label{eq:dimless}
\begin{split}
    &\hat \lambda_\alpha^2= \frac{\Omega_d\Gamma_\alpha }{Z_\alpha^{3} \Lambda^{2-d}} \lambda_\alpha^2, \quad
    \hat \lambda_\theta^2=\frac{ \Omega_dZ_\alpha^{1/3}\Gamma_\theta^2}{ \Gamma_\alpha Z_{d}^{10/3} \Lambda^{2-d}} \lambda_\theta^2,\\
    &\quad\hat g=\sqrt{ \frac{\Omega_d\Gamma_\alpha}{Z_\alpha^{5/3} Z_{d}^{4/3}\Lambda^{2-d}} } g, \quad r=\frac{Z_d}{Z_\alpha},
    \end{split}
\end{equation}
to absorb all bare and anomalous dimensions. Here $r$ is the weak scaling parameter, which flows to $0$ (or infinity) if the wavefunction renormalizations $Z_d$ and $Z_\alpha$ of $\theta$ and $\alpha$ pick up different scalings, leading to the weak scaling anticipated above.\\
The scaling exponents $z_{\alpha,\theta}$ and $\chi_{\alpha,\theta}$ as defined in \eqref{eq:2pointCorr} are related to the wavefunction renormalizations via
\begin{align}
    z_{\alpha,\theta}&=2-\nu_{\alpha,\theta},\\
    2\chi_{\alpha,\theta}&=2-d-\nu_{\alpha,\theta}+\eta_{\alpha,\theta},
\end{align}
where
\begin{align}
    \nu_{\alpha,\theta}&=-\frac{\Lambda\partial_\Lambda Z_{\alpha,\theta}}{Z_{\alpha,\theta}},\\
    \eta_{\alpha,\theta}&=-\frac{\Lambda\partial_\Lambda \Gamma_{\alpha,\theta}}{\Gamma_{\alpha,\theta}}.
\end{align}
In the rotating phase, we only find weak coupling fixed points of dimensionless flow equations with $r\rightarrow 0$. In this limit, the flow equations are
\begin{subequations}\label{eq:rotating_flow}

\begin{align}
    \nu_d&\equiv-\Lambda\partial_{\Lambda} Z_d=0,\\
    \nu_c&\equiv-\Lambda\partial_{\Lambda} Z_c=0,\\
    \eta_\theta &=0,\\
    \nu_\alpha&=\eta_\alpha=\hat{\lambda}_\alpha^2,\\
    \beta_{\hat{\lambda}_\alpha}&=\frac{1}{2}\left(d-2-\eta_\alpha+3\nu_\alpha\right)\hat{\lambda}_{\alpha}+4(N-2)\hat{\lambda}_\theta\hat{g''}^2,\\
    \beta_{\hat{\lambda}_\theta}&=\frac{1}{2}\left(d-2+\eta_\alpha-2\eta_\theta+10/3\nu_d-1/3\nu_\alpha\right)\hat{\lambda}_\theta,\\
    \beta_{\hat{g}}&=\frac{1}{2}\left(d-2-\eta_\alpha+5/3\nu_\alpha+4/3\nu_d\right)\hat{g}.
\end{align}
\end{subequations}
First, we see that below two dimensions, the Gaussian fixed point is not stable and the system always flows to a 'rough' strongly interacting fixed point. 
In $2-\epsilon$ dimensions,
\begin{align}
    \nu_\alpha=\eta_\alpha =\frac{3\epsilon}{2}
\end{align}
and the fixed point couplings are also of $O(\epsilon)$. This yields the following dynamical and roughness exponents:
\begin{equation}\label{eq:rot_exponents}
\begin{split}
    &z_\theta=2,\quad 2\chi_\theta=2\chi_\alpha=\epsilon,\\
    &z_\alpha=2-\frac{3\epsilon}{2}.
    \end{split}
\end{equation}
Above two dimensions, i.e. for $\epsilon<0$ the Gaussian fixed point is stable and the interacting fixed point has one relevant direction. It thus corresponds to the critical point separating a smooth regime with Gaussian fluctuations from a rough regime. This is fully analogous to the qualitative phase diagram of the KPZ equation above and below two dimensions. \\
The key takeaway is that in low dimensions, the system always flows towards a rough and weak scaling fixed point, characterized by the two differing dynamical exponents in Eq.~\eqref{eq:rot_exponents}. This is thus expected to be observable in the rotating phase in $d=1,2$ without additional fine tuning. Going to $d=1$ -- i.e. setting $\epsilon=1$ -- lies outside the regime where the present perturbative expansion is valid. However, we expect the qualitative features (rough and weak scaling) to transcend into the nonperturbative regime, and we confirm this through numerical simulations in Sec. \ref{sec:num_sim}. Furthermore, if the bare couplings are small, the length scales at which they become nonperturbative, i.e. the length scale $\xi_i$ where $\hat\lambda_i(\xi_i)\sim O(1)$ in terms of the bare coefficients of the model that are defined at some UV scale $a\sim \Lambda^{-1}$, $\hat{\lambda}_{\alpha,0},\hat{\lambda}_{\theta,0},\hat{g}_{0}$ 
\begin{align}
    \xi_\alpha=\frac{a}{\hat\lambda_{\alpha,0}^2},\,\xi_\theta=\frac{a}{\hat\lambda_{\theta,0}^2},\,\xi_g=\frac{a}{\hat{g}_0^2}.
\end{align}
In a two-dimensional system, all couplings are marginally relevant, as in the standard KPZ case \cite{Kardar1986}.
Therefore, for weak coupling strengths $|\lambda|$ the nonequilibrium behavior emerges above scales that are algebraically large in $|\lambda|^{-1}$ and in two dimensions above an exponentially large scale. To evade this exponential scale in a finite size experiment, one has to operate in a regime where the nonlinearity is large already at the microscopic scale.

\subsection{Flow equations and scaling in the oscillating phase}

The same analysis can be repeated in the oscillating phase. There, we recover the stochastic dynamical system studied in \cite{Ertas1992,Ertas1993}. The resulting renormalization group equations are reproduced and we review the results here. The dimensionful flow equations of the oscillating phase can be straightforwardly obtained from the ones of the rotating phase by setting $Z_c=g''=0$ and replacing the prefactors of $2N-4$ that arise from traces over the complex $\theta$-fields with the $N-1$ for the oscillating case. The key difference with respect to the rotating phase is that because $g\in\mathbb{R}$ or $g''=0$ the dimensionful flow of $\lambda_\alpha$, Eq.~\eqref{eq:KPZ_rotating_dimless} vanishes, like in the KPZ case. The \emph{dimensionless} flow equations then display the peculiar feature that $\beta_{\lambda_i}\propto\hat{\lambda}_i$. Therefore the relative signs of $\lambda_\alpha,\lambda_\theta$ and $g$ cannot change during the flow. The flow separates in three distinct blocks depending on the signs of the three couplings $\lambda_\alpha,\lambda_\theta$ and $g$. If $\lambda_\theta$ and $g$ have the same sign, the system flows to a KPZ fixed point for $N\leq5$. For  $N\geq5$ the system flows to weak scaling fixed points that have been observed numerically in the original works \cite{Ertas1992,Ertas1993}. By introducing the weak scaling parameter $r$, one can also see this analytically in the flow equations \cite{zelle2025Diss}. This procedure further allows one to access weak scaling fixed points for different combinations of the coupling signs.
Numerical simulations \cite{Ertas1993} suggest that this regime is unstable and heats up. 

In conclusion, the oscillating phase only exists when the nonlinearities share the same signe. Within the oscillating phase  one observes the KPZ universality class. Differing signs of the nonlinearities lead to fluctuations that immediately destroy coherence on short distances and are therefore incompatible with the time-crystalline phase. 

\section{Numerical observation of universal scaling in one dimension}\label{sec:num_sim}

The one-loop RG presented above predicts universally rough and weakly scaling correlations for the Goldstone modes in one dimension in the rotating time-crystalline phases of the $O(N)\times SO(2)$. In one dimension there is no true long range order: Due to strong Goldstone fluctuations, correlation functions asymptotically decay to zero. Here we will quantify the precise degree of this decay. It is of stretched-exponential type, with exponents revealing the scaling behavior analysed above. Resolving such behavior  has lead to the successful observation of the ordinary KPZ scaling function in the decay of correlation functions in  one- and two-dimensional exciton-polariton condensates \cite{Fontaine2022,Widmann2026}.\\
We can access these correlation functions numerically by using the equivalence of the dynamical action of the $O(N)\times SO(2)$ model \eqref{eq:action} to a Langevin equation,
\begin{align}\label{eq:GPE}
    \partial_t\vecpsi_c+\frac{\delta H_d}{\delta\vecpsi_c^\dagger}+i\frac{\delta H_c}{\delta\vecpsi_c^\dagger}+\vecxi=0
\end{align}
with $H_{c,d}$ as defined in \eqref{eq:nonhermitian_hamiltonian} and $\vecxi$ a Gaussian white noise with $\langle\vecxi^\dagger(t,\vecr)\vecxi(t',\vecr')\rangle=2\Gamma_0\delta(t-t')\delta(\vecr-\vecr')$. This equation of motion can be simulated directly on one-dimensional chains and the respective correlation functions can be computed.
For the numerical simulations, we discretized space and time and expressed all quantities in lattice units, such that the couplings in Eq.~\eqref{eq:nonhermitian_hamiltonian} are effectively dimensionless.
We simulate the full Langevin equation~\eqref{eq:GPE}, rather than the effective models
\eqref{eq:rot_goldstone_action} and \eqref{eq:osc_goldstone_action}, in order to
preserve compactness at the numerical level. 
Details of the numerical implementation are provided in Appendix~\ref{app:numerical}.
Goldstone fluctuations around the mean field solution $\psi_{0,A}$ of the rotating phase lead to a correlator
\begin{equation}
\begin{split} C_\psi(t,\vecr)=&\langle\vecpsi^\dagger(t_0+t,\vecr_0+\vecr)\vecpsi(t_0,\vecr_0)\rangle \\
=&\rho_0\langle e^{i(\alpha(t_0+t,\vecr_0+\vecr)-\alpha(t_0,\vecr_0))}\\
&\cdot e^{\sum_{a,i}\theta_{a,i}(t_0+t,\vecr_0+\vecr)T^{(a,i)}-\sum_{a,i}\theta_{a,i}(t_0,\vecr_0)T^{(a,i)}}\rangle.
\end{split}
\end{equation}
In a cumulant expansion this yields, to leading order,
\begin{align}
    C_\psi(t,\vecr)\approx\rho_0 e^{-\frac{1}{2} \mathcal{C}_\alpha(t,\vecr)-\frac{1}{2}\mathcal{C}_\theta(t,\vecr)},
\end{align}
with $\mathcal{C}_\alpha(t,\vecr)=\langle\left(\alpha(t_0+t_,\vecr_0+\vecr)-\alpha(t_0,\vecr_0)\right)^2\rangle$ and $\mathcal{C}_\theta(t,\vecr)=\langle\left(\theta(t_0+t_,\vecr_0+\vecr)-\theta(t_0,\vecr_0)\right)^2\rangle$.
Since we predicted $\mathcal{C}_\alpha$ to grow more slowly than $\mathcal{C}_\theta$ in the rotating phase, the exponential decay of the correlation function will be sensitive to the universal scaling function of the chronon mode $\alpha$
\begin{align}
    -\ln{C_\psi(t,\vecr)}\sim \mathcal{C}_\alpha(t,\vecr).
\end{align}
To distill the scaling of $\theta$ fluctuations, we consider another object, $\phi_\theta(t,\vecr):=|\psi_1(t,\vecr)|^2$. The choice of the first component is arbitrary. Importantly, this object is invariant under the $SO(2)$ rotations generated by $\alpha$ but remains sensitive to $\theta$ fluctuations that include the first component $\phi_\theta(t,\vecr)\approx \rho_0\cos^2\theta$. Therefore, in a cumulant expansion, we get
\begin{equation}  
\begin{split}
-\ln& \langle \phi_\theta (t_0+t,\vecr_0+\vecr) \phi_\theta (t_0,\vecr_0)\rangle \\
    &\sim \sum_{a,i} b_{a,i}\langle \left( \theta_{a,i}(t_0+t,\vecr_0+\vecr) + \theta_{a,i}(t_0,\vecr_0) \right)^2 \rangle \\
    &\sim \mathcal C_\theta(t,\vecr),
\end{split}
\end{equation}
where $b_{a,i}$ are a set of numbers arising from the relative direction between the first axis and the plane in which the system rotates.
\\
This second observable is easily available from numerics, but its physical interpretation is less transparent. In view of such interpretation, we focus on the most relevant case $N=3$, and consider the angular momentum of the rotation that fixes the rotational plane. Its fluctuations are dominated by reorientations of the plane, i.e. the $\theta$ modes. In terms of the complex order parameter field it is given by
\begin{align}
    \boldsymbol{L}(t,\vecr)=\operatorname{Re} \vecpsi(t,\vecr)\times \operatorname{Im} \vecpsi(t,\vecr).
\end{align}
If we neglect massive fluctuations of the amplitude, we can rewrite this as
\begin{equation}
    \boldsymbol{L}(t,\vecr) =\rho_0 e^{\sum_{a,i}\theta_{a,i}(t,\vecr)T^{(a,i)} }\hat{e}_3 .
\end{equation}
Its correlation function follows the same scaling law as $\phi_\theta$,
\begin{align}
    -\ln \langle \boldsymbol{L} (t_0+t,\vecr_0+\vecr) \boldsymbol{L}(t_0,\vecr_0)\rangle\sim \mathcal{C}_\theta(t,\vecr).
\end{align}
Since the $\phi_\theta$ object is numerically more easily accessible, we do not use the angular momentum to extract the scaling exponents.
\begin{figure}[t]
    \centering
    \includegraphics[width=1\linewidth]{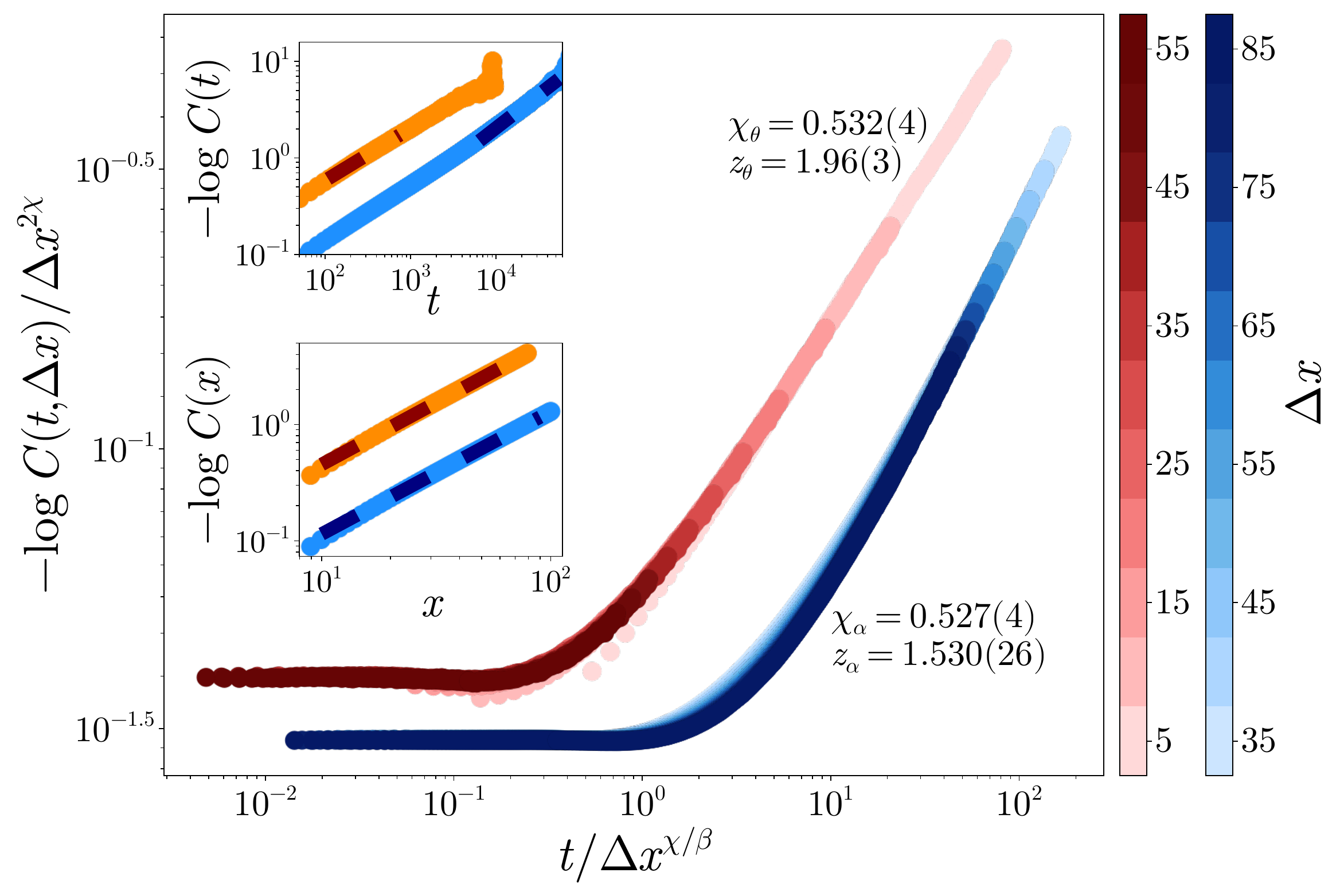}

    \caption{Scaling results of the functions $C_\psi$ (blue) and $C_\theta$ (red) in the rotating phase for $N=3$. 
    Extracted exponents are listed in Table \ref{tab:scalingExponentsOverview}.
    The main panel shows temporal correlations for spatial separations $\Delta x$, measured in units of the lattice spacing, in steps of $5$.
    The collapse exponents are obtained from the insets: the upper inset shows the time correlation at $\Delta x=0$, the lower inset the equal-time correlation. We observe weak scaling, here for couplings $Z_d=1$, $Z_c=-1$, $r_d=-1$, $u_d=1$, $\kappa_d=0.5$, $u_c=-0.2$, $\kappa_c=-0.5$, $r_c=-0.3$, and  $\Gamma_0=0.035$ with a system size $L=1000$. The average is over 15,801 trajectories.}
    \label{fig:rot_phase_N=3_gc=0.2}
\end{figure}

\begin{figure}[t]
    \centering
    \includegraphics[width=1\linewidth]{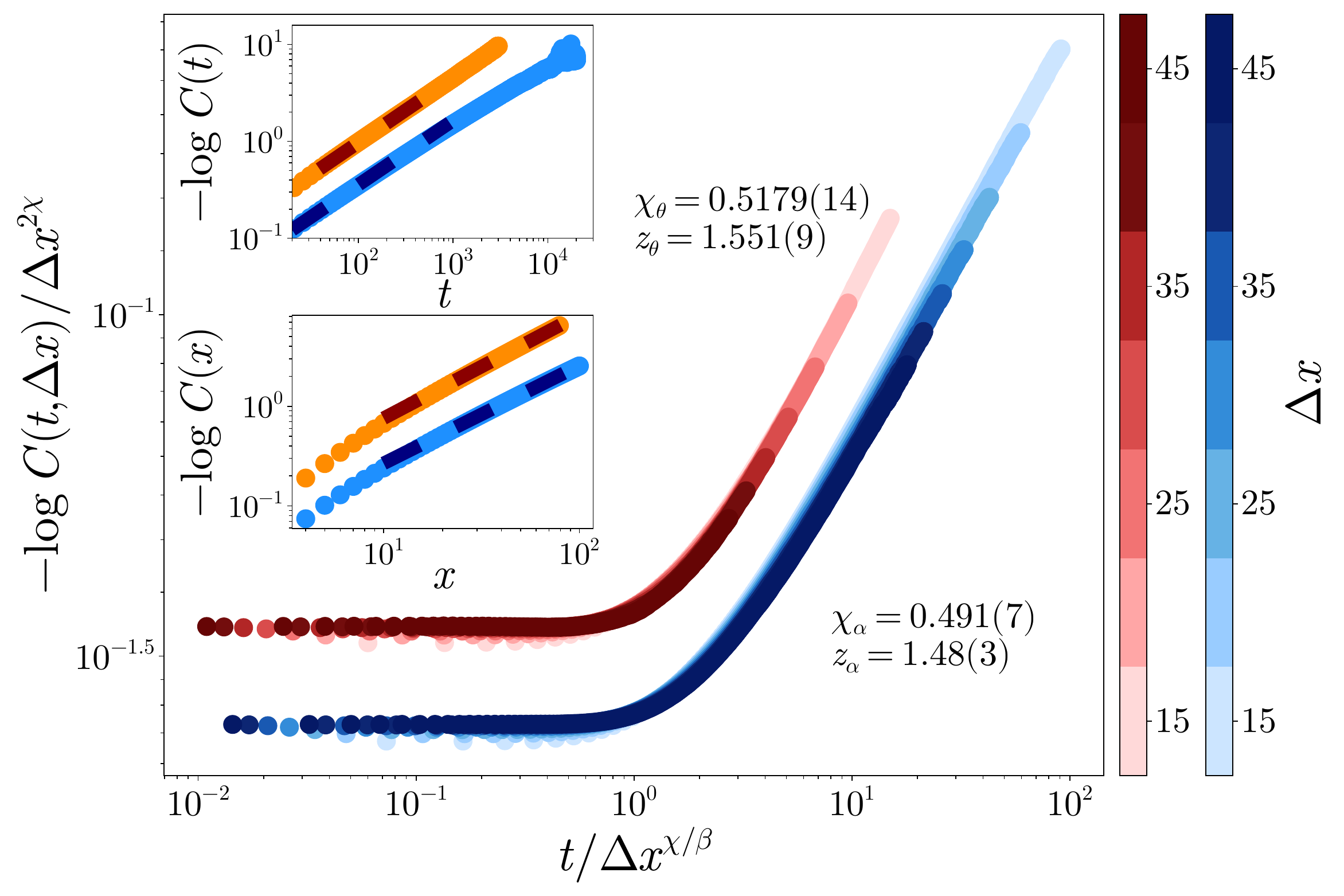}
    
    \caption{
Scaling results of the functions $C_\alpha$ (blue) and $C_\theta$ (red) in the oscillating phase for $N=3$. 
The presentation and extraction procedure are analogous to Fig.~\ref{fig:rot_phase_N=3_gc=0.2}; extracted exponents are listed in Table~\ref{tab:scalingExponentsOverview}. 
Couplings are $Z_d=1$, $Z_c=0$, $r_d=-1$, $u_d=1$, $\kappa_d=-0.25$, $u_c=1.6$, $\kappa_c=-0.4$, $r_c=-2.1$, and $\Gamma_0=0.02$ in a system with $L=10,000$ sites. 
The average is over 9,501 trajectories.}
    \label{fig:osc_phase_N=3_KPZ}
\end{figure}

\subsection{Scaling in the rotating phase}

For the analysis in this section, we choose the couplings such that the nonlinear terms, $\lambda_\alpha$ and $\lambda_\theta$ in equation (\ref{eq:rot_goldstone_action}), are as large as possible without entering the regime in which topological defects affect the correlations.
These defects emerge for sufficiently large $\lambda_\alpha$ and will be discussed in more detail in Sec. \ref{sec:TopDef}.
In Fig. \ref{fig:rot_phase_N=3_gc=0.2}, the scaling of the correlation functions $C_\psi$ and $C_\theta$ is shown.
The scaling exponents are identified via the asymptotics 
\begin{equation}
\begin{split}
    \mathcal{C}_{\alpha,\theta}(t,0)&\propto t^{2\beta_{\alpha,\theta}}, \,\beta_{\alpha,\theta}=\frac{\chi_{\alpha,\theta}}{z_{\alpha,\beta}},\\
    \mathcal{C}_{\alpha,\theta}(0,\vecr)&\propto r^{2\chi_{\alpha,\theta}}.
\end{split}
\end{equation}
For sufficiently large time scales, the correlation functions exhibit weak scaling in the equal-space correlations, implying that $\beta_\alpha > \beta_\theta$.
This is visible in the scaling collapse, where at large times the slope is steeper for $C_\psi$ than for $C_\theta$.
The numerical values of these exponents, obtained from a fit, are summarized in table \ref{tab:scalingExponentsOverview}.

The equal-time correlation functions of the \(\alpha\) and \(\theta\) modes exhibit similar scaling to each other, with a scaling exponent of \(\chi_{\alpha,\theta} \approx 0.53\), which is slightly larger than the KPZ value \(\chi = 1/2\).

The corresponding dynamical exponents are given by the ratio between $\chi$ and $\beta$.
We note that the exponent pairs $(\beta_\alpha,\chi_\alpha)=(0.345,0.527)$ and $(\beta_\theta,\chi_\theta)=(0.2706,0.532)$ are close to, yet clearly distinct from, the KPZ and EW values, respectively.
Therefore, a decoupling into a KPZ mode and a diffusive EW mode can be excluded.
This is consistent with the RG analysis, according to which the coupling between the $\alpha$ and $\theta$ fluctuations is relevant.
Using these exponents, Fig. \ref{fig:rot_phase_N=3_gc=0.2} shows that there is a full scaling collapse of the space-time dependent correlators.
Furthermore, scaling exponents obtained from numerical simulations are universal in that they do not depend on the couplings chosen for the numerics as we demonstrate in the appendix \ref{app:numerical}.

This confirms the field-theoretic prediction that time-crystalline order in non-Abelian symmetry groups can indeed give rise to weak scaling behavior with universal scaling exponents.


\subsection{Scaling in the oscillating phase}\label{subsec:numOsc}

We now perform the same analysis in the oscillating phase. 
We choose couplings corresponding to the scenario in which all nonlinear coupling terms in the effective equation (\ref{eq:osc_goldstone_action}) have the same sign and the field theory thus predicts KPZ scaling behavior for both $\mathcal{C}_\theta$ and $\mathcal{C}_\alpha$. 
Furthermore, in the oscillating phase we can define $\phi_\alpha(t,x):=\vecpsi(t,x)^T \vecpsi(t,x)$.

This object has the convenient property of being independent of the $\vectheta$ modes in the oscillating phase. 
The $\vectheta$ fluctuations act as local $O(N)$ rotations of the field configuration and can be written as $\vecpsi(t,x)=R(\vectheta)\vecvarphi(t,x)$, where $\vecvarphi(t,x)$ contains the phase fluctuation $\alpha$ as well as possible massive fluctuations. 
Since $R(\vectheta)$ is an element of $O(N)$, the rotation matrix satisfies $ R(\vectheta)^T R(\vectheta)=\mathbb{1}$. 
Therefore, $\phi_\alpha(t,x)=\vecpsi^{\,T}(t,x)\vecpsi(t,x)=\vecvarphi^{\,T}(t,x)\vecvarphi(t,x)$, showing that the $\theta$ modes cancel out.

Fig.~\ref{fig:osc_phase_N=3_KPZ} shows the correlation functions $C_\alpha$ and $C_\theta$ in the oscillating phase.
By fitting the equal-time and equal-space correlation functions, we identify a scaling regime in which the simulated exponents match the KPZ prediction very well, $z_{\alpha,\theta}\approx 3/2$ and $\chi_{\alpha,\theta}\approx 1/2$. 
We conclude that the scaling of the correlations of $\alpha$ and $\theta$ in the analyzed regime is consistent with the KPZ universality class.

\begin{table}
    \centering
    \begin{tabular}{|c|c|c|}
    \hline
     & Oscillating phase & Rotating phase \\
    \hline
    $\beta_\alpha$ & $0.3323(29)$ & $0.345(7)$ \\
    \hline
    $\chi_\alpha$  & $0.491(7)$ & $0.527(4)$\\
    \hline
    $z_\alpha$  & $1.48(3)$ & $1.530(26)$ \\
    \hline
    $\beta_\theta$  & $0.3338(18)$ & $0.2706(29)$ \\
    \hline
    $\chi_\theta$  & $0.5179(14)$ & $0.532(4)$  \\
    \hline
    $z_\theta$  & $1.551(9)$ & $1.96(3)$ \\
    \hline
    \end{tabular}
    \caption{Scaling exponents of the $\alpha$ and $\theta$ fields in the oscillating and rotating phase. The $\beta$ and $\chi$ exponents are measured through a fit and the $z$ exponents is given by the ratio $\chi/\beta$ for $N=3$.}
    \label{tab:scalingExponentsOverview}
\end{table}

\section{Topological defects}\label{sec:TopDef}
\begin{figure}[t]
    \centering
    \includegraphics[width=1\linewidth]{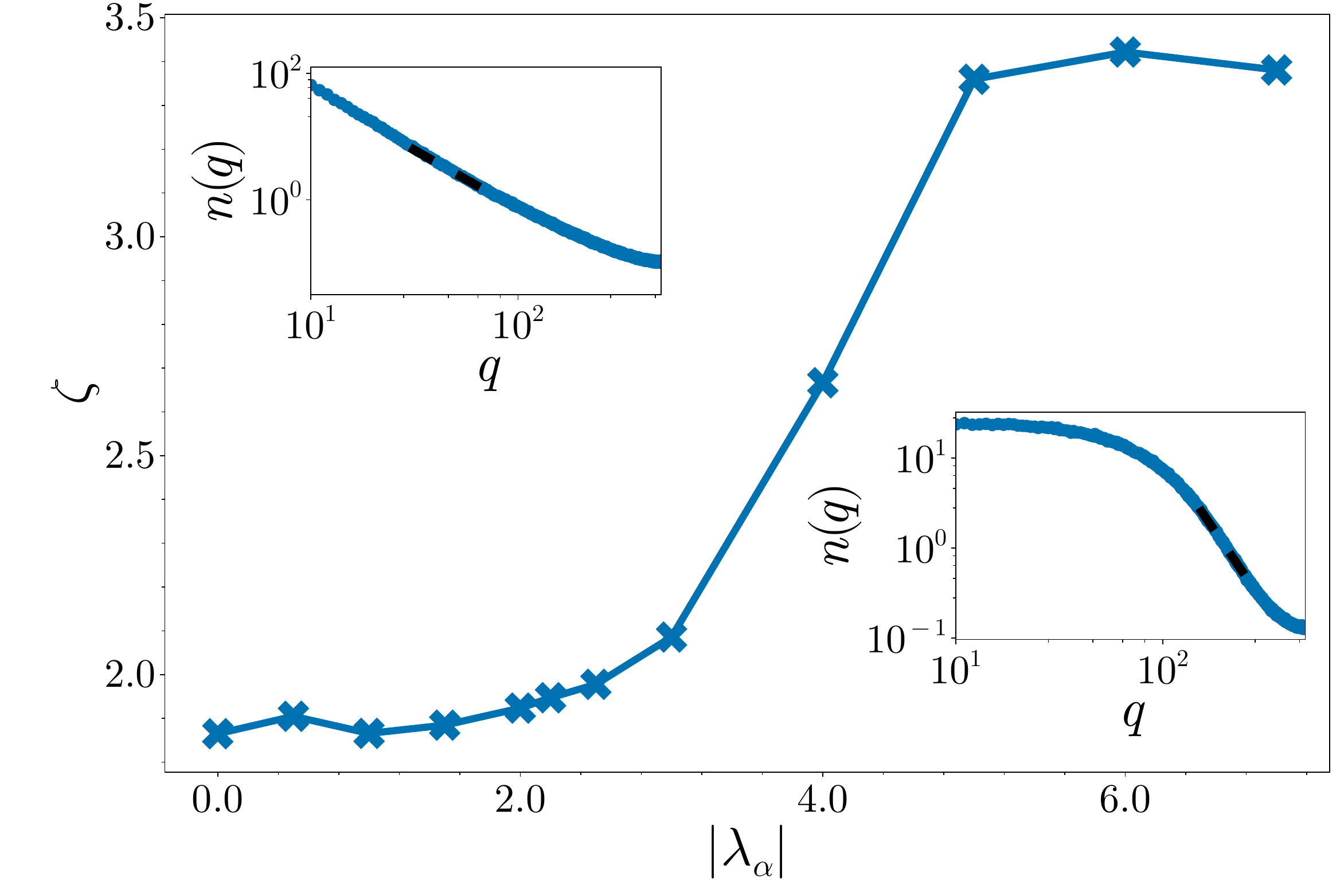}
    \caption{
In the rotating phase, the momentum distributions exhibit scaling behavior $n(q)\propto q^{-\zeta}$ for large $q$, for different coupling ratios $\lambda_\alpha$ with $\Gamma_0=0.05$, for $N=3$. The exponent of the momentum distribution exhibits a crossover from a defect-free regime to a regime dominated by defects. The left inset shows the form of the momentum distributions for $|\lambda_\alpha|=0.0$. The right inset shows the form of the momentum distributions for $|\lambda_\alpha|=7.0$.
    The used couplings are $Z_d=1$, $Z_c=-1$, $r_d=-1$, $u_d=1$, $\kappa_d=0.5$, $\kappa_c=-0.5$ and $r_c=-0.3$. $\lambda_\alpha$ is tuned by changing $u_c$ from $-1.0$ to $+2.5$.}
    \label{fig:mom_dis_rot_overview_plus_inserts}
\end{figure}

\begin{figure}[t]
    \centering
    \includegraphics[width=1\linewidth]{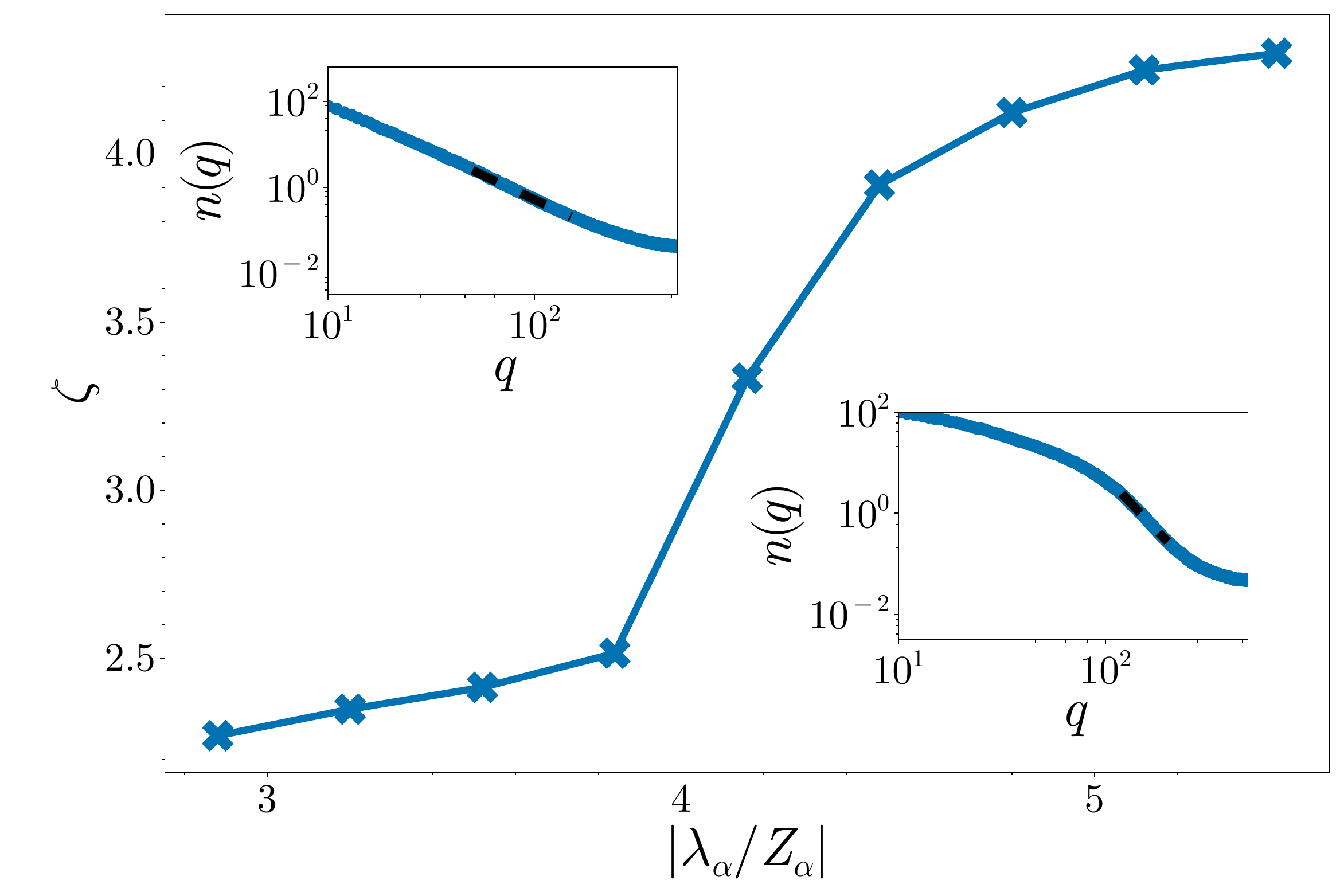}
    \caption{
    In the oscillating phase, the momentum distributions exhibit scaling behavior $n(q)\propto q^{-\zeta}$ for large $q$, for different coupling ratios $|\lambda_\alpha/Z_\alpha|$ with $\Gamma_0=0.02$, for $N=3$. The left inset shows the form of the momentum distributions for $|\lambda_\alpha/Z_\alpha|=2.88$. The right inset shows the form of the momentum distributions for $|\lambda_\alpha/Z_\alpha|=5.44$.
    The used couplings are $Z_d=1$, $Z_c=0$, $r_d=-1$, $u_d=1$, $\kappa_d=-0.25$.
    To tune $|\lambda_\alpha/Z_\alpha|$, we set $u_c=0.8x$ and $\kappa_c=-0.2x$ and change $x$ from $1.8$ to $3.4$.}
    \label{fig:mom_dis_first_q_overview_plus_inserts}
\end{figure}

We have seen that the flat approximation \eqref{eq:rot_goldstone_action} of the nonlinear sigma model action \eqref{eq:nlsm_rot} correctly predicts that the  equilibrium-breaking interactions of the Goldstone modes give rise to strongly interacting, weak scaling behavior. This approximation however misses an important aspect: the Goldstone modes live on curved, compact manifolds, and thus topological defect configurations, such as vortices, are possible.\\
Here we embark on a first --  and very rough -- investigation of the nature, the regimes of appearance, and the physical impact of these defects. 
\\ 
Some orientation can be gleaned from the case of a single compact KPZ Goldstone mode, i.e. $N=1$. The defects are vortices, which are expected to unbind on asymptotic length scales \cite{Sieberer2016a, Wachtel2016, He2015, Daviet2025}, where they then destroy KPZ scaling. They are, however, exponentially suppressed at low noise and for low coupling strengths. Numerical analysis of the compact KPZ equation in $1+1$ dimensions shows that at low noise, there is a nonequilibrium transition at a critical strength of the nonlinearity into a regime where vortices proliferate, a phenomenon dubbed 'vortex turbulence'~\cite{He2017, Vercesi_2024}. We now analyze to what extent such scenario carries over to the larger symmetry groups considered here.

We begin by determining the point-like topological space-time excitations in $1+1$ dimensions that can arise in the cosets describing the non-Abelian time-crystals. They can be classified by mappings from closed loops in coordinate space into the coset space~\cite{Mermin1979, Chaikin1995}. 
Two such mappings are said to be homotopic if they can be continuously deformed into one another.
In contrast, mappings that differ by the presence of a topological defect, such as a vortex, are not homotopic. 
The first homotopy group $\pi_1(G/H)$ of the coset space endows these mappings with a group structure, which also governs the composition of defects.

\begin{figure}
    \includegraphics[width=0.9\linewidth]{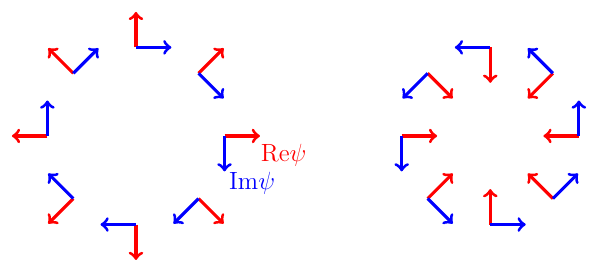}
    \caption{Vortex and anti-vortex for $N=2$ in the rotating phase. For $N=3$, these two configurations can be in fact continuously transformed into each other by performing rotations along the directions of the imaginary and real part of the order parameter. }
    \label{fig:topDef}
\end{figure}

In the rotating phase, the coset that hosts the order parameter is $\mathcal{M}^{\mathrm{rot}}=O(N)\times SO(2)/O(N-2)\times SO(2)\simeq O(N)/ O(N-2)$. The topological defects for this symmetry group have been explored, for example, in the context of frustrated magnets~\cite{Kawamura1984,Kawamura1998}.
For \(N=2\), the order-parameter manifold is \(O(2)\), whose connected components satisfy $\pi_1\bigl(O(2)\bigr) \cong \mathbb{Z}$.
The corresponding topological defects are vortices characterized by an integer winding number.
For \(N=3\), the coset space is isomorphic to \(SO(3)\) and $\pi_1\bigl(SO(3)\bigr) \cong \mathbb{Z}_2$.
Consequently, there is a single non-trivial homotopy class of topological defects. 
Fig.~\ref{fig:topDef} illustrates configurations with winding numbers \(+1\) and \(-1\). 
Although these configurations are distinct for \(N=2\), they belong to the same non-trivial homotopy class for \(N=3\) and can therefore be continuously deformed into one another.
For \(N>3\), one has $\pi_1\bigl(O(N)/O(N-2)\bigr)=0$, so there are no topologically stable vortex defects.
To analyze defects that are not point-like in $1+1$ dimensions, we need to consider the second homotopy group. 
It gives the group structure for maps from closed surfaces in coordinate space to the coset space. 
In the rotating phase, $\pi_2(O(N)/O(N-2))\cong\mathbb{Z}$ if $N=4$ \cite{Hoo1965}.

In the oscillating phase, the order-parameter manifold is locally parameterized by $SO(2)\times O(N)/O(N-1)\simeq S^1\times S^{N-1}$.
This local parameterization, however, does not capture the global diagonal $\mathbb{Z}_{2,d}$ identification of the order parameter. Indeed, a phase rotation by $\pi$ can be compensated for by an inversion of the $O(N)$ order parameter.
The order-parameter manifold is therefore
\begin{equation}
    \mathcal{M}^{\mathrm{osc}}
    = \frac{S^1\times S^{N-1}}{\mathbb{Z}_{2,\mathrm{d}}}.
\end{equation}
The vortex classification depends on $N$.
For $N=1$, we find $\pi_1\left(\mathcal{M}^{\mathrm{osc}}\right)=\mathbb{Z}$ with the ordinary $2\pi$ phase vortex~\cite{He2017}.
For $N=2$, the manifold is a torus with $\pi_1(\mathcal{M}^{\mathrm{osc}})\cong\mathbb{Z}\times\mathbb{Z}$, generated by combined half-quantum vortices in which the phase and the $O(2)$ angle wind simultaneously, either in the same or in opposite directions.
For $N>2$, $\pi_1(\mathcal{M}^{\mathrm{osc}})\cong\mathbb{Z}$, but its generator is a half-quantum vortex consisting of a $\pi$ phase vortex together with an inversion of the $O(N)$ order parameter~\cite{Kawaguchi2012}.
Consequently, an ordinary $2\pi$ phase vortex carries topological charge two.
The second homotopy group of the oscillating order-parameter manifold is
$\pi_2(\mathcal{M}^{\mathrm{osc}})\cong\mathbb{Z}$ for $N=3$, allowing for topologically stable skyrmion textures.

\begin{table}
    \centering
    \begin{tabular}{c|c|c}
    & Oscillating phase &Rotating phase \\ 
    Coset space & $S^1\times S^{N-1}/\mathbb{Z}_{2,\mathrm{d}}$& $O(N)/O(N-2)$ \\ 
    \hline \hline
    $N=1$ & $\mathbb{Z}$ & / \\
    $N=2$ & $\mathbb{Z} \times \mathbb{Z}$ & $\mathbb{Z}$ \\ 
    $N=3$ & $\mathbb{Z}$&$\mathbb{Z}_2$ \\ 
    $N>4$ & $\mathbb{Z}$&$0$
    \end{tabular}
    \caption{First homotopy groups $\pi_1$ of the coset space hosting the order parameter for oscillating and rotating phases.
    }
    \label{fig:homotopy_class}
\end{table}

We then study whether -- in analogy to the $N=1$ case --  at a fixed, finite system size there is a nonequilibrium transition or crossover from a regime where the nontrivial weak scaling dominates, into a regime where defects proliferate as one increases the coupling strength at a fixed, low noise level. The onset of such a vortex dominated regime can be  detected numerically through the momentum dependence of the autocorrelation function $n(q)=\langle \vecpsi^*(q)\cdot \vecpsi(q)\rangle$.
In the phase with few vortices, its  behavior at large momenta is expected to be $n(q)\propto q^{-2}$.\\
Before turning to the $1+1$ dimensional case, consider a static two-dimensional system with uncorrelated vortices~\cite{Schmidt_2012}.  The field configuration of a single vortex obeys $|\nabla \vecpsi(\mathbf r)| \sim 1/r$, which implies the scaling $|\nabla \vecpsi(\mathbf q)|^2 \sim q^{-2}$ in momentum space. On the other hand, the same quantity can be expressed in terms of the momentum distribution as $|\nabla \vecpsi(\mathbf q)|^2 \sim q^2 n(q)$. 
Matching both expressions yields the scaling behavior
\begin{equation}
    n(q) \sim q^{-4}.
\end{equation}
Since the dynamical critical exponent $z\neq1$ we cannot immediately apply this result to the $1+1$ dimensional case. 
Moreover, vortices in the system are not isolated but can overlap and interact, which is expected to lead to deviations from this simple scaling argument. 
Nevertheless, the momentum distribution in the vortex-dominated regime is still expected to follow $n(q)\propto q^{-\zeta}$ with $\zeta > 2$. This behavior is also observed at the vortex proliferation threshold for the $N=1$ compact KPZ equation in~\cite{He2017}.\\
Figs.~\ref{fig:mom_dis_rot_overview_plus_inserts} and \ref{fig:mom_dis_first_q_overview_plus_inserts} show the scaling form of the momentum distributions in the rotating and oscillating phases from numerical simulations.

In both cases, a crossover is observed as the nonlinear coupling $|\lambda_\alpha|$ is increased in the respective phase.
For sufficiently small $|\lambda_\alpha|$, we observe $\zeta \sim 2$, which can be associated with the vortex-free regime.
The scaling behavior changes for larger nonlinear couplings and $\zeta$ approaches $\zeta \approx 3.5$ in the rotating phase and $\zeta \approx 4.4$ in the oscillating phase.
These scaling exponents are consistent with our expectations for a regime dominated by vortices.
We therefore conclude that both the rotating and oscillating phases exhibit a crossover from a regime in which vortices are rare to a regime in which they dominate the system dynamics. 
This crossover, analyzed here in the low-noise regime, is driven entirely by the strength of the nonlinear couplings in the respective phases.

\section{Discussion and Outlook}

We have developed the nonlinear sigma model framework for intrinsically nonequilibrium phases of matter, which describe wide classes of time crystals that break non-Abelian symmetries. We identified novel universal scaling regimes that go far beyond the KPZ paradigm, both quantitatively and qualitatively, and sharply distinguish these nonthermal phases from phenomena accessible within purely Hamiltonian dynamics in low dimensions. Importantly, this nontrivial scaling emerges already for arbitrarily weak violations of equilibrium conditions: nonequilibrium acts as a relevant perturbation that destabilizes diffusive behavior at long wavelengths. We confirmed this scenario through direct numerical simulations of models  
within the $O(N)\times SO(2)$ universality class.

The framework developed here opens a new arena for nonequilibrium statistical mechanics, while at the same time providing concrete guidance for the understanding and design of state-of-the-art solid state experiments.

To begin with the latter, several new avenues emanate from the present work. For example, microscopic anisotropies and disorder may substantially enrich the structure of these phases. Furthermore, inclusion of additional hydrodynamic modes associated with conserved quantities is relevant for many  active and condensed matter scenarios, where conserved densities generically participate in the dynamics. Finally, incorporating couplings to gauge fields — either through the electromagnetic environment or as emergent degrees of freedom in quantum materials — will provide valuable probes of universality in linear response~\cite{Diessel2025SteadyStates,Zelle2026SC}, and may even reveal qualitatively new universal gauge phenomena far from equilibrium due to the intrinsically massless nature of these modes~\cite{Coleman1973,Halperin1974}. 

From the perspective of statistical mechanics, the framework established here opens a broad range of future directions. The most immediate interpretation views our construction as a direct generalization of KPZ physics from a single real scalar degree of freedom to multiple interacting modes with a symmetry structure inherited from the underlying coset, while retaining the characteristic universal flow structure as a function of dimension. One is then naturally led to the same central questions that define the original KPZ problem: What are the precise values of the critical exponents and universal scaling functions in different dimensions? Does the theory possess an upper critical dimension? How do the rich symmetry structures and Ward identities known for KPZ generalize to these non-Abelian settings? For the NLSMs describing rotating phases, these questions are entirely new; for oscillating phases, they acquire renewed significance in view of the rapidly expanding experimental opportunities. These problems could be addressed numerically, or within the approximation neglecting curvature discussed in Sec.~\ref{sec:rg} using powerful functional renormalization group techniques~\cite{Canet2010KPZ,Canet2011KPZ,Kloss2012KPZ,Dupuis2021FRGReview}. 

Further key directions connect more directly to aspects of NLSMs without an analogue in the original KPZ problem: curvature and topology. 
One central question concerns the interplay between nonequilibrium interactions and the phase transitions of nonlinear sigma models in higher dimensions. Perturbative analysis in $4-\epsilon$ dimensions already reveals genuinely nonthermal universality classes emerging at transitions of $O(N)\times SO(2)$ models~\cite{Daviet2024}; another particularly interesting expansion point is $2+\epsilon$ dimensions, where both the equilibrium NLSM and the original KPZ problem exhibit phase transitions, albeit of fundamentally different character. Finally, the topology of the cosets permits the formation of a variety of real-space defects. Here we obtained a first glimpse of their impact by focusing on $SO(2)$ vortex defects, which both control the asymptotic scaling behavior and can drive new topological transitions. The role of more complex defects in the phase structure of $SO(2)\times O(N)$ symmetric models remains a fascinating and largely unexplored problem.

\acknowledgements

We thank Alexander Altland, Luca Delacretaz, Ruchira Mishra and Martin Zirnbauer for valuable discussions. This research was supported by Deutsche Forschungsgemeinschaft (DFG, German Research Foundation) through CRC1238 project C04 number 277146847 and CRC183 project C01 number 277101999. CZ was supported by DFG through project number 570906600.
Our numerical simulations were performed on the RAMSES cluster at RRZK Cologne.

\FloatBarrier
\newpage
\onecolumngrid
\appendix

 \section{One-loop integrals}\label{app:loops}

In this appendix, we give the derivations of the one-loop dimensionful RG $\beta$-functions. These were first published in the first authors' dissertation \cite{zelle2025Diss}. 
In phase A, the self energy contributions the retarded Green's functions read
\begin{align}
    &\Sigma^R_{\alpha}(\omega=0,\vecp)=
    \lambda_\alpha^2\int_{\vecq,\omega}\vecq\cdot(\vecq-\vecp)\vecp\cdot(\vecq-\vecp)G^R_\alpha(\omega,\vecq)G^K_\alpha(\omega,\vecq-\vecp)
    +\vecq\cdot(\vecp-\vecq)\vecq\cdot\vecp G^R_\alpha(\omega,\vecq-\vecp)G^K_\alpha(\omega,\vecq)\\ \nonumber
    &+(N-2)\Big[\lambda_\theta (g'+ig'')\int_{\vecq,\omega}\vecq\cdot(\vecq-\vecp)\vecp\cdot(\vecq-\vecp)G^R_\theta(\omega,\vecq)G^K_\theta(\omega,\vecq-\vecp)
   +\vecq\cdot(\vecp-\vecq)\vecq\cdot\vecp G^R_\theta(\omega,\vecq-\vecp)G^K_\theta(\omega,\vecq)+c.c\Big]
\end{align}
and 
\begin{align}
\nonumber
    &\Sigma^R_{\theta}(\omega=0,\vecp)=(g'+ig'')^2\int_{\vecq,\omega}\vecq\cdot(\vecq-\vecp)\vecp\cdot(\vecq-\vecp)G^R_\theta(\omega,\vecq)G^K_\alpha(\omega,\vecq-\vecp)+\vecq\cdot(\vecp-\vecq)\,\vecq\cdot\vecp G^R_\theta(\omega,\vecq-\vecp)G^K_\alpha(\omega,\vecq)\\ 
    &+\lambda_\theta (g'+ig'')\int_{\vecq,\omega}\vecq\cdot(\vecq-\vecp)\vecp\cdot(\vecq-\vecp)G^R_\alpha(\omega,\vecq)G^K_\theta(\omega,\vecq-\vecp)+\vecq\cdot(\vecp-\vecq)\vecq\cdot\vecp G^R_\alpha(\omega,\vecq-\vecp)G^K_\theta(\omega,\vecq).
\end{align}
The noise contributions read
\begin{align}
\Sigma^K_\alpha(0,0)=&\frac{1}{2}\lambda_\alpha^2\int_{\vecq,\omega}\vecq^4G^K_\alpha(\omega,\vecq)^2+\frac{1}{2}\lambda_\theta^2(N-1)\int_{\vecq,\omega}\vecq^4G^K_\theta(\omega,\vecq)^2,\\
\Sigma^K_\theta(0,0)=&({g'}^2+{g''}^2)\int_{\vecq,\omega}\vecq^4G^K_\alpha(\omega,\vecq)G^K_\theta(\omega,\vecq),
\end{align}

where the Green's functions are
\begin{align}
    G^R_{\alpha}(\omega,\vecq)&=\frac{1}{-i\omega+Z_{\alpha}\vecq^2}, \quad G^R_{\theta}(\omega,\vecq)&=\frac{1}{-i\omega+(Z_d+iZ_c)\vecq^2},\quad G^K_{\theta,\alpha}(\omega,\vecq)=-2\Gamma_{\theta,\alpha}|G^R_{\theta,\alpha}(\omega,\vecq)|^2.
\end{align}
We can perform the frequency integrations analytically and project the momentum dependence of the spectral self energy and arrive at
\begin{align}
    \partial_{\vecp^2}\Sigma_{\alpha}^R(0,0) =& \frac{(d-2)\left(2\gamma_\alpha Z_d^2\lambda_\alpha^2+(N-2)\gamma_\theta Z_\alpha^2g'\lambda_\theta\right)}{d Z_\alpha^2Z_d^2}\Omega_d\int_{0}^\Lambda dq\,q^{d-3}\\
    \partial_{\vecp^2}\Sigma_{\theta}^R(0,0) =&\Big(\frac{g^2\gamma_\alpha(4Z_\alpha-d(Z_\alpha+Z_\theta))}{dZ_\alpha(Z_\alpha+Z_\theta)^2}-\frac{2\gamma g\lambda_\theta(Z_\alpha+Z_\theta^*-4Z_d)}{Z_d(Z_\alpha+Z_\theta^*)^2}\Big)\Omega_d\int_{0}^\Lambda dq\,q^{d-3}\\
  \Sigma^K_\alpha(0,0)=&\Big(\frac{\gamma_\alpha^2\lambda_\alpha}{Z_\alpha^3}+2(N-2)\frac{\gamma_\theta^2\lambda_\theta}{Z_d^3}\Big)\Omega_d\int_{0}^\Lambda dq\,q^{d-3}\\
    \Sigma^K_\alpha(0,0)=&\frac{|g|^2\gamma_\alpha\gamma_\theta(Z_\alpha+Z_d)}{Z_\alpha Z_d(Z_c^2+(Z_\alpha+Z_d)^2)}\Omega_d\int_{0}^\Lambda dq\,q^{d-3}.
\end{align}
Here, $\Omega_d$ is the area of the sphere in $d$ dimensions divided by $(2\pi)^d$ and $Z_\theta=Z_d+iZ_c,\, g=g'+ig''$.

\begin{figure}
    \centering
    \subfloat[Retarded self energy contributions to $\alpha$]{
        \includegraphics[width=0.585\linewidth]{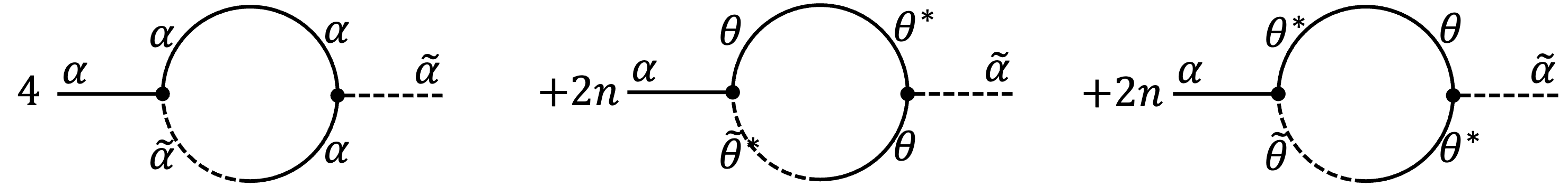}}\hfil 
    \subfloat[Retarded self energy contributions to $\theta$]{
        \includegraphics[width=0.39\linewidth]{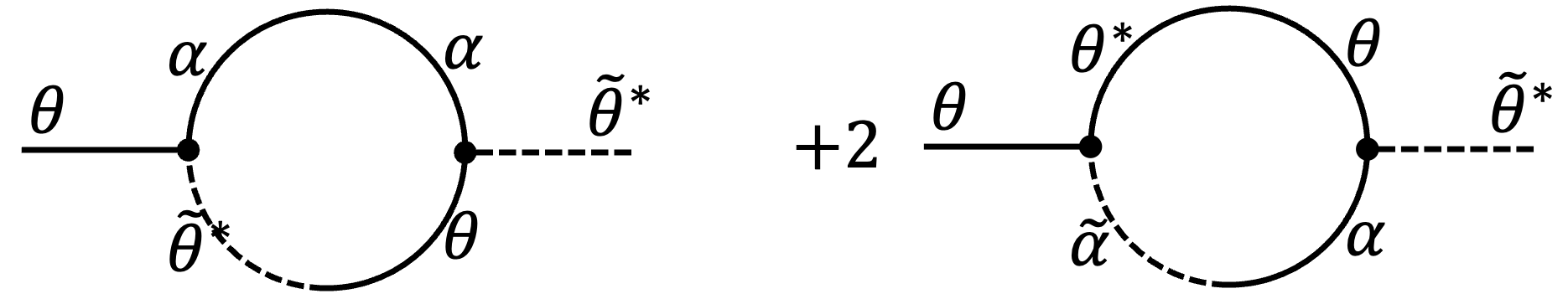}}\\
    \subfloat[Noise contributions to $\gamma_\alpha$]{
        \includegraphics[width=0.39\linewidth]{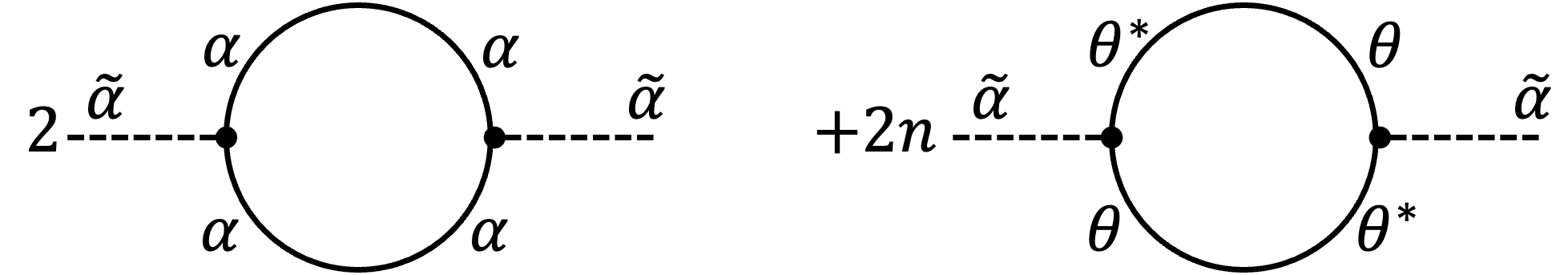}}\hfil
    \subfloat[Noise contributions to $\gamma_\theta$]{
       $\quad\quad$ \includegraphics[width=0.1625\linewidth]{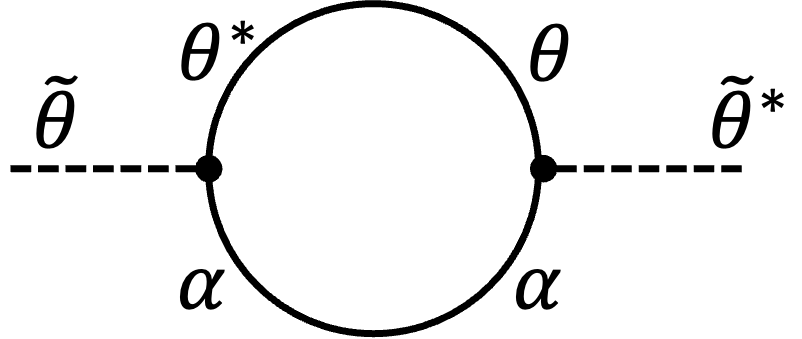}$\quad\quad$}
    \caption{Diagrammatic one-loop contributions to the self-energies.}
    \label{fig:rotating_propagator_diagrams}
\end{figure}

\begin{figure}
    \centering
    \subfloat[Contributions to $\lambda_\alpha$]{
        \includegraphics[width=\linewidth]{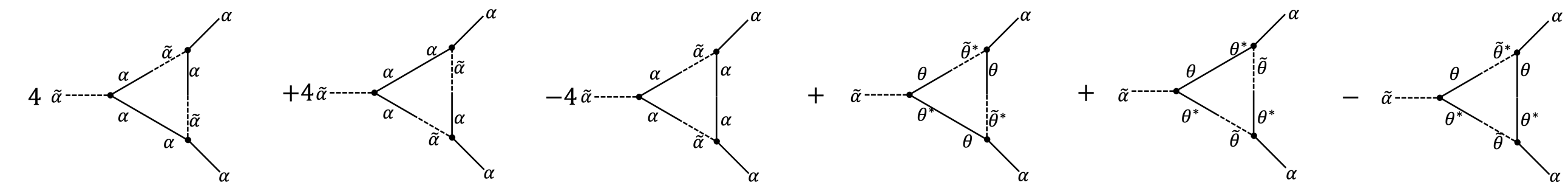}}\\
    \subfloat[Contributions to $\lambda_\theta$]{
        \includegraphics[width=\linewidth]{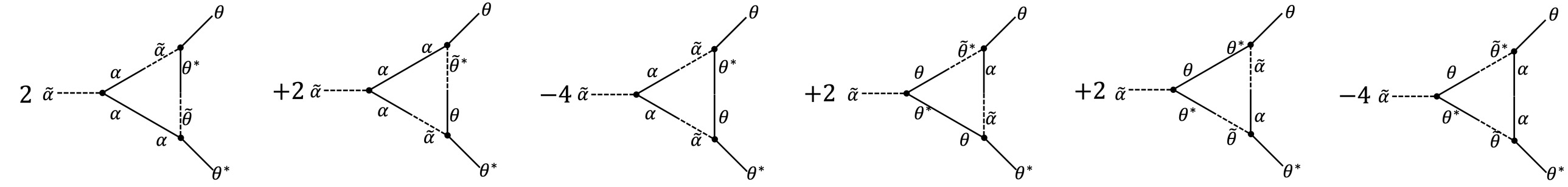}}\\
\subfloat[Contributions to $g'-ig''$]{
        \includegraphics[width=\linewidth]{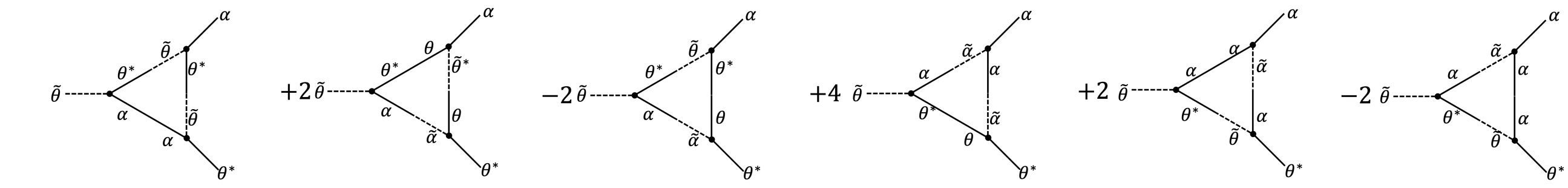}}

    \caption{Diagrammatic one-loop contributions to the couplings}
    \label{fig:rotating_coupling_diagrams}
\end{figure}

 We used $\int\dd^dq (\vecq\cdot\vecp)^2f(\vecq^2)=(p/d)^2\int \dd^dq\vecq^2f(\vecq^2)$. These self energies are the one-loop corrections of $Z_{\theta,\alpha},2\Gamma_{\theta,\alpha}$. From this, we can derive the \emph{dimensionful} $\beta$-functions given in the main text by simply taking a derivative with respect to the UV cutoff $\Lambda$. This is equivalent to a Wilsonian momentum shell RG with UV sharp cutoffs. 

In an analogous approach, we derive the dimensionful $\beta$-functions of phase couplings from the vertex corrections by evaluating the Feynman diagrams in Fig.~\ref{fig:rotating_coupling_diagrams}:
\begin{align}
    \Lambda\partial_\Lambda Z_\alpha =& \Lambda^{d-2}\frac{(d-2)\left(2\gamma_\alpha Z_d^2\lambda_\alpha^2+(N-2)\gamma_\theta Z_\alpha^2g'\lambda_\theta\right)}{d Z_\alpha^2Z_d^2}\\
    \Lambda\partial_\Lambda Z_\theta =&\Lambda^{d-2}\Big(\frac{g^2\gamma_\alpha(4Z_\alpha-d(Z_\alpha+Z_\theta))}{dZ_\alpha(Z_\alpha+Z_\theta)^2}-\frac{2\gamma_\theta g\lambda_\theta(Z_\alpha+Z_\theta^*-4Z_d)}{Z_d(Z_\alpha+Z_\theta^*)^2}\Big)\\
   \Lambda\partial_\Lambda\gamma_\alpha=&\Lambda^{d-2}\Big(\frac{\gamma_\alpha^2\lambda_\alpha^2}{Z_\alpha^3}+2(N-2)\frac{\gamma_\theta^2\lambda_\theta\lambda_\alpha}{Z_d^3}\Big)\\
    \Lambda\partial_\Lambda\gamma_\theta=&\Lambda^{d-2}\frac{|g|^2\gamma_\alpha\gamma_\theta(Z_\alpha+Z_d)}{Z_\alpha Z_d(Z_c^2+(Z_\alpha+Z_d)^2)}\\
    \Lambda\partial_\Lambda\lambda_\alpha=&\Lambda^{d-2}\frac{4(N-2){g''}^2\lambda_\theta}{Z_d^3}\\ 
    \Lambda\partial_\Lambda\lambda_\theta=&\frac{\lambda_\theta\Lambda^{d-2}}{Z_\alpha^2Z_d^2(Z_c^2+(Z_\alpha+Z_d)^2)}\Bigg(2\gamma_\alpha Z_d\Big(2\lambda_\alpha Z_d(g''Z_c+g'(Z_\alpha+Z_d))-|g|^2Z_\alpha(Z_\alpha+Z_d)\Big)\\ \nonumber
    &+4\gamma_\theta Z_\alpha \lambda_\theta\Big(g' Z_{\alpha}(Z_\alpha+Z_d)-g''Z_\alpha Z_c -2\lambda_\alpha Z_d(Z_\alpha+Z_d)\Big)\Bigg)\\ 
    \Lambda\partial_\Lambda g=&\Lambda^{d-2} g \Bigg(\frac{2g\gamma_\alpha(2\lambda_\alpha-g)}{dZ_\alpha(Z_\alpha+Z_\theta)^2}-4\gamma_\theta\lambda_\theta\frac{(Z_\alpha+Z_\theta^*)^2(g'Z_d+ig''(Z_\alpha+Z_d+Z_\theta))+2Z_d(Z_\alpha+Z_\theta)^2\lambda_\alpha}{dZ_d^2(Z_c^2+(Z_d+Z_\alpha)^2)^2}\Bigg).
\end{align}

We now introduce dimensionless couplings to find the RG fixed points underlying the scale invariance of the system:
\begin{equation}
\begin{split}
    &\hat \lambda_\alpha^2= \frac{\Omega_d\Gamma_\alpha }{Z_\alpha^{3} \Lambda^{2-d}} \lambda_\alpha^2, \quad
    \hat \lambda_\theta^2=\frac{ \Omega_dZ_\alpha^{1/3}\Gamma_\theta^2}{ \Gamma_\alpha Z_{d}^{10/3} \Lambda^{2-d}} \lambda_\theta^2,\\
    &\quad\hat g=\sqrt{ \frac{\Omega_d\Gamma_\alpha}{Z_\alpha^{5/3} Z_{d}^{4/3}\Lambda^{2-d}} } g, \quad r=\frac{Z_d}{Z_\alpha}, \quad r_K=\frac{Z_d}{Z_c}
    \end{split}
\end{equation}

The weak scaling fixed point described can be found as following: $\Lambda\partial_\Lambda r=r(\operatorname{\nu_\theta}-\nu_\alpha)$ and therefore a fixed point with $r=0$ is stable if $\operatorname{Re}{\nu_\theta}<\nu_\alpha$ at $r=0$. In the present case, in $d=1$ at $r=0$ $\nu_\theta=0$ and $\nu_\alpha=\hat{\lambda}_\alpha^2$ so the stability condition is satisfied. Then, using the fixed point value $r=0$ and the dimensionless couplings defined as above yields the remaining flow equations given in the main text Eq.~\eqref{eq:rotating_flow}.

\section{Obtaining Goldstone mode effective action}\label{app:phase_amplitude}

In the following section the effective actions for our Goldstone modes are derived explicitly from the initial action \ref{eq:action}. This allows effective couplings to be extracted from our coefficients for initial action. Were we to add extra terms to our Hamiltonian \ref{eq:nonhermitian_hamiltonian}, we would expect different coefficients. However, as mentioned in the main text, the general form of the NLSM actions \ref{eq:nlsm_rot}, \ref{eq:nlsm_osc} are complete and no extra terms could be added given internal and space-time symmetry constrains.

For the following we begin with the Hamiltonian for the full model:

\begin{equation}
     H_{c,d} = \int_{\vecr}\boldsymbol{\psi}^*\cdot(-Z_{c,d}\nabla^2 + \gamma_{c,d})\boldsymbol{\psi} + \frac{u_{c,d}}{4} \rho^2 + \frac{\kappa_{c,d}}{4}\tau,
\end{equation}
with $\rho=\vecpsi^*\cdot\vecpsi$ and $\tau=\left(\vecpsi\cdot\vecpsi\right)\left(\vecpsi^*\cdot\vecpsi^*\right)$. 
\subsection{Rotating Phase}
For the rotating phase  we have the condition that $\kappa_d<0$ giving us the solution that $\text{Re}\, \vecpsi_0 \perp \text{Im} \,\vecpsi_0$. Solving for the mean field solution: $\rho_0=\frac{-2\gamma_d}{u_d}$. Note that we can always include a phase $\exp{i\omega_0 t}$
such that $\omega_0=-\gamma_c-\frac{u_c}{2}\rho_0$.

Without loss of generality we chose $\vecpsi_0 = e^{i\omega_0 t}\sqrt{\rho_0}\mathbf{x}$ where $\mathbf{x}=\frac{\hat{\mathbf{e}}_1+i\hat{\mathbf{e}}_2}{2}$ and expand the saddle point solution: $\vecpsi=\sqrt{\rho_0+\delta\rho}R(\alpha,\boldsymbol{\theta})\left(\mathbf{x}+(\delta\sigma_1+i\delta\sigma_2)\mathbf{x}^*\right)$ where $\delta\sigma_i \ll1$, and $\frac{\delta\rho}{\rho_0}\ll1$. We also redefine $\vecpsi_q=R(\alpha,\boldsymbol{\theta})\left((\chi_1+i\chi_3)\mathbf{x}+(\chi_2+i\chi_4)\mathbf{x}^*+\boldsymbol{\xi}_1+i\boldsymbol{\xi}_2\right)$. Note that $\boldsymbol{\xi}_i\cdot \hat{\mathbf{e}}_1=\boldsymbol{\xi}_i\cdot \hat{\mathbf{e}}_2=0$. Here we identify our $2N-3$ Goldstone modes $\alpha$ and $\boldsymbol{\theta}$. We see that the remaining degrees of freedom are massive and can be integrated out with their conjugate field. We use an adiabatic approximation resulting in the following action:

\begin{equation}
\begin{aligned}
    S=\int_{t,\vecr}&\sqrt{\rho_0}\left((\chi_1-i\chi_3)\mathbf{x}^\dagger+(\chi_2-i\chi_4)\mathbf{x}+\boldsymbol{\xi}^T_1-i\boldsymbol{\xi}^T_2\right)R^T\left(\partial_t-Z\nabla^2\right)R\mathbf{x}\\&+\rho_0^\frac{3}{2}\left(\kappa(\chi_2-i\chi_4)(\delta\sigma_1+i\delta\sigma_2)+\frac{u}{2}\frac{\delta\rho}{\rho_0}(\chi_1-i\chi_3)\right)+c.c. +2i\gamma\left(\chi_i\chi^i+\boldsymbol{\xi}^T_1\boldsymbol{\xi}_1+\boldsymbol{\xi}^T_2\boldsymbol{\xi}_2\right).
\end{aligned}
\end{equation}
After integration of massive modes and conjugate pairs $\left(\delta\rho,\delta\sigma_1,\delta\sigma_2,\chi_1,\chi_2,\chi_4\right)$ we have the Goldstone effective action:

\begin{equation}
\begin{aligned}
    S=\int_{t,\vecr}&-2i\sqrt{\rho_0}\chi_3\mathbf{x}^\dagger\left(R^T\partial_tR-Z_\alpha\nabla(R^T\nabla R)+i\lambda_\alpha\nabla R^T \nabla R\right)\mathbf{x}\\&+\sqrt{\rho_0}(\boldsymbol{\xi}^T_1-i\boldsymbol{\xi}^T_2)\left(R^T\partial R_t-Z_\theta\nabla(R^T\nabla R)-ig\nabla R^T \nabla R\right)
    \mathbf{x}+c.c.\\& +2i\Gamma_\alpha\chi_3^2+2i\Gamma\left(\boldsymbol{\xi}^T_1\boldsymbol{\xi}_1+\boldsymbol{\xi}^T_2\boldsymbol{\xi}_2\right),
\end{aligned}
\end{equation}
with coefficients:
\begin{equation}
    \begin{aligned}
        Z_\alpha&=Z_d\left(1+ \frac{Z_c}{Z_d} \frac{u_c}{u_d}\right)\\
        Z_\theta&=Z_d+iZ_c\\
        \lambda_{\alpha}&=\lambda_\theta= Z_c\left(1 - \frac{Z_d}{Z_c} \frac{u_c}{u_d} \right) \\
        g&=iZ_d-Z_c\\
        \Gamma_\alpha&=\Gamma\left(1+\left(\frac{u_c}{u_d}\right)^2\right).
    \end{aligned}
\end{equation}
\subsection{Oscillating phase} 
For the oscillating phase we have the condition that $\kappa_d>0$ giving us the solution that $\text{Re} \,\vecpsi_0 \parallel \text{Im} \,\vecpsi_0$. Solving for the mean field solution: $\rho_0=\frac{-2\gamma_d}{u_d+\kappa_d}$. Note that we can always include a phase $\exp{i\omega_0 t}$
such that $\omega_0=-\gamma_c-\frac{u_c+\kappa_c}{2}\rho_0$. Without loss of generality we chose $\vecpsi_0 = \sqrt{\rho_0}e^{i\omega_0 t}\hat{\mathbf{e}}_1$ and expand the saddle point solution: $\vecpsi=e^{i\alpha}\sqrt{\rho_0+\delta\rho}R(\boldsymbol{\theta})\left(\hat{\mathbf{e}}_1+i\vec{\delta\sigma}\right)$ where $\hat{\mathbf{e}}_1\cdot\vec{\delta\sigma}=0$, and $|\vec{\delta\sigma}| \ll1$, $\frac{\delta\rho}{\rho_0}\ll1$. We also redefine $\vecpsi_q=e^{i\alpha}R(\boldsymbol{\theta})\left(\boldsymbol{\xi}_1+i\boldsymbol{\xi}_2\right)$. Here we can associate our $N$ degrees of freedom $\alpha$ and $\boldsymbol{\theta}$ with our $N$ Goldstone modes. We see that the remaining degrees of freedom are massive and can be integrated out with their conjugate field. We use an adiabatic approximation resulting in the following action:

\begin{equation}
\begin{aligned}
    S=\int_{t,\vecr}&\sqrt{\rho_0}e^{-i\alpha}\left(\boldsymbol{\xi}^T_1-i\boldsymbol{\xi}^T_2\right)R^T(\boldsymbol{\theta})\left(\partial_t-Z\nabla^2\right)e^{i\alpha}R(\boldsymbol{\theta})\hat{\mathbf{e}}_1\\&+\rho_0^\frac{3}{2}\left(\boldsymbol{\xi}^T_1-i\boldsymbol{\xi}^T_2\right)\left(\frac{\delta\rho}{\rho_0}\frac{u+\kappa}{2}\hat{\mathbf{e}}_1-i\kappa\vec{\delta\sigma}\right)+c.c. +2i\Gamma\left(\boldsymbol{\xi}^T_1\boldsymbol{\xi}_1+\boldsymbol{\xi}^T_2\boldsymbol{\xi}_2\right).
\end{aligned}
\end{equation}

After integration of the massive modes and conjugate pairs $\left(\xi_2^i,\delta\rho^i\right)$ for $i\neq1$ and $\left(\xi_1^1,\delta\rho\right)$ we are left with the action:

\begin{equation}
\begin{aligned}
    S=\int_{t,\vecr}&2\sqrt{\rho_0}\xi_2^1\left(\partial_t\alpha-Z_\alpha\nabla^2\alpha+\lambda_\alpha(\nabla\alpha)^2+\lambda_\theta\nabla R^T\nabla R\right)+2i\Gamma (\xi_2^1)^2\\&+2\sqrt{\rho_0}\boldsymbol{\xi}^T_1R^T\left(\partial_t-Z_\theta\nabla^2+g\nabla \alpha \nabla \right)R\hat{\mathbf{e}}_1 +2i\Gamma_\theta\boldsymbol{\xi}^T_1\boldsymbol{\xi}_1,
\end{aligned}
\end{equation}
where we note that now $\boldsymbol{\xi}_1\cdot\hat{\mathbf{e}}_1=0$ and the couplings are: 
\begin{equation}
    \begin{aligned}
        Z_\alpha&=Z_d\left(1+ \frac{Z_c}{Z_d} \frac{u_c+\kappa_c}{u_d+\kappa_d}\right)\\
        Z_\theta&= Z_d\left(1+ \frac{Z_c}{Z_d} \frac{u_c}{u_d}\right)\\
        \lambda_{\alpha}&=\lambda_\theta= Z_c\left(1 - \frac{Z_d}{Z_c} \frac{u_c+\kappa_c}{u_d+\kappa_d} \right) \\
        g&=2Z_c \left(1 - \frac{Z_d}{Z_c} \frac{\kappa_c}{\kappa_d} \right)\\
        \Gamma_\theta&=\Gamma\left(1+\left(\frac{\kappa_c}{\kappa_d}\right)^2\right)\\
        \Gamma_\alpha&=\Gamma\left(1+\left(\frac{\kappa_c+u_c}{\kappa_d+u_d}\right)^2\right).
    \end{aligned}
\end{equation}

\section{Gaussian stationary distributions in $d=1$ dimensions}\label{app:fokker_planck}

In the following, we demonstrate how the model admits a stationary Gaussian probability distribution, for values other than just the trivial the non-interacting diffusive equilibrium problem, by mapping to the Fokker Planck equation. We see a similar behavior in the single-Goldstone-mode KPZ equation \cite{Kamenev2023}.

For the generic Langevin equations for complex fields $\phi^\mu$ and response fields $\tilde{\phi}^\mu$,
\begin{equation}
    \partial_t\phi^\mu=A^\mu(\phi)+D^{\mu\nu}(\phi)\tilde{\phi}^{\nu},
\end{equation}
the associated Fokker-Planck equation reads 
\begin{equation}
    \partial_t\mathcal{P}[\phi;t]=-\partial_\mu J^\mu;\hspace{5mm} \partial_\mu=\frac{\delta}{\delta\phi^\mu};\hspace{5mm} J^\mu=A^\mu(\phi)\mathcal{P}[\phi;t]-\partial_\nu(D^{\mu\nu}(\phi)\mathcal{P}[\phi;t]).
\end{equation}

\subsection{Rotating Phase}

From the Langevin equations
\begin{equation}
    \begin{aligned}
        \partial_t\alpha=&Z_\alpha\nabla^2\alpha-\lambda_\alpha (\nabla\alpha)^2 -\lambda_\theta|\nabla\boldsymbol{\theta}|^2 -i\Gamma_\alpha\tilde{\alpha}\\
        \partial_t\boldsymbol{\theta}=&Z_\theta\nabla^2\boldsymbol{\theta}-g \nabla\alpha \nabla\boldsymbol{\theta}  -i\Gamma_\theta\tilde{\boldsymbol{\theta}}\\
        \partial_t\boldsymbol{\theta}^*=&Z_\theta\nabla^2\boldsymbol{\theta}^*-g \nabla\alpha \nabla\boldsymbol{\theta}^*  -i\Gamma_\theta\tilde{\boldsymbol{\theta}}^*,
    \end{aligned}
\end{equation}
we find the Fokker-Planck equation
\begin{equation}
    \partial_t\mathcal{P}[\phi;t]=-\partial_\mu J^\mu;\hspace{5mm} \partial_\mu=\frac{\delta}{\delta\phi^\mu};\hspace{5mm} \phi^\mu=(\alpha,\vectheta^T,\vectheta^\dagger)^\mu,
\end{equation}
where 
\begin{equation}
    \begin{aligned}
        J^\alpha&=(Z_\alpha\nabla^2\alpha-\lambda_\alpha (\nabla\alpha)^2 -\lambda_\theta|\nabla\boldsymbol{\theta}|^2)\mathcal{P}[\phi;t]-\Gamma_\alpha\frac{\delta\mathcal{P}[\phi;t]}{\delta\alpha},\\
        J^{\vectheta}&=(Z_\theta\nabla^2\boldsymbol{\theta}-g \nabla\alpha \nabla\boldsymbol{\theta} )\mathcal{P}[\phi;t]-\Gamma_\theta\frac{\delta\mathcal{P}[\phi;t]}{\delta\vectheta^*},\\
        J^{\vectheta^*}&=(Z_\theta^*\nabla^2\vectheta^*-g^* \nabla\alpha \nabla\vectheta^* )\mathcal{P}[\phi;t]-\Gamma_\theta\frac{\delta\mathcal{P}[\phi;t]}{\delta\vectheta}.
    \end{aligned}  
\end{equation}
Using an ansatz in terms of a Gaussian stationary probability distribution,
\begin{equation}
    \mathcal{P}[\boldsymbol{\theta},\boldsymbol{\theta^*},\alpha] \propto \exp \left(\int_{\vecr}-\frac{Z_d}{\Gamma_\theta}|\nabla\boldsymbol{\theta}|^2-\frac{Z_\alpha}{2\Gamma_\alpha}(\nabla\alpha)^2\right).
\end{equation}
We will see that a Gaussian stationary distributions holds for specific parameter choices. 
First we study the current arising from the imaginary (dissipative) terms in the Langevin equation:
\begin{equation}
J^{\boldsymbol{\theta}}_Z=iZ_c\nabla^2\boldsymbol{\theta}\mathcal{P};\hspace{3mm}J^{\boldsymbol{\theta^*}}_Z =-iZ_c\nabla^2\boldsymbol{\theta^*}\mathcal{P}
\end{equation}
which has vanishing divergence irrespective of the couplings,
\begin{equation}
    \partial_{\vectheta} J^{\vectheta}+\partial_{\vectheta^*} J^{\vectheta^*}=0.
\end{equation}

Second, we consider currents arising from the nonlinearities (KPZ-type terms), where we get constraints on the parameters:
\begin{equation}
\begin{aligned}
    J_{\text{KPZ}}^\alpha &=(\lambda_\alpha (\nabla\alpha)^2 +\lambda_\theta|\nabla\boldsymbol{\theta}|^2)\mathcal{P},\\
    J_{\text{KPZ}}^{\boldsymbol{\theta}}&=g\nabla\alpha\nabla\boldsymbol{\theta}\mathcal{P},\\
    J_{\text{KPZ}}^{\boldsymbol{\theta^*}}&=g^*\nabla\alpha\nabla\boldsymbol{\theta^*}\mathcal{P}.
\end{aligned}
\end{equation}
In $d=1$ we can identify a total derivative,
\begin{eqnarray}
\partial_i\partial^i\phi\partial_j\phi\partial^j\phi\propto\partial_i(\partial^i\phi\partial_j\phi\partial^j\phi); \hspace{3mm} \partial_i=\frac{\partial}{\partial x^i},
\end{eqnarray}
where $\partial_i$ denotes a real space derivative.

As such we see that we require:
\begin{equation}
    \int_x\frac{\lambda_\theta Z_\alpha}{\Gamma_\alpha}|\nabla\boldsymbol{\theta}|^2\nabla^2\alpha+\frac{gZ_d}{\Gamma_\theta}\nabla\alpha\nabla\boldsymbol{\theta}\nabla^2\boldsymbol{\theta^*}+\frac{g^*Z_d}{\Gamma_\theta}\nabla\alpha\nabla\boldsymbol{\theta^*}\nabla^2\boldsymbol{\theta}=0,
\end{equation}
which holds if $\mathrm{Im}(g)=0$ and $\frac{\lambda_\theta Z_\alpha}{\Gamma_\alpha}=\frac{\mathrm{Re}(g)Z_d}{\Gamma_\theta}$. Additionally $\lambda_\alpha\in\mathbb{R}$ is compatible with these conditions in $d=1$.

\subsection{Oscillating Phase}
 We repeat the same derivation as above, now noting that all the fields are real; these results also appear in  \cite{Ertas1993}.
From the Langevin equations
\begin{equation}
    \begin{aligned}
        \partial_t\alpha=&Z_\alpha\nabla^2\alpha-\lambda_\alpha (\nabla\alpha)^2 -\lambda_\theta(\nabla\vectheta)^2 -i\Gamma_\alpha\tilde{\alpha}\\
        \partial_t\boldsymbol{\theta}=&Z_\theta\nabla^2\boldsymbol{\theta}-g \nabla\alpha \nabla\boldsymbol{\theta}  -i\Gamma_\theta\tilde{\boldsymbol{\theta}},
    \end{aligned}
\end{equation}
we find the Fokker-Planck equation
\begin{equation}
    \partial_t\mathcal{P}[\phi;t]=-\partial_\mu J^\mu;\hspace{5mm} \partial_\mu=\frac{\delta}{\delta\phi^\mu};\hspace{5mm} \phi^\mu=(\alpha,\vectheta^T)^\mu , 
\end{equation}
where 
\begin{equation}
    \begin{aligned}
        J^\alpha&=(Z_\alpha\nabla^2\alpha-\lambda_\alpha (\nabla\alpha)^2 -\lambda_\theta(\nabla\vectheta)^2\mathcal{P}[\phi;t]-\Gamma_\alpha\frac{\delta\mathcal{P}[\phi;t]}{\delta\alpha},\\
        J^{\vectheta}&=(Z_\theta\nabla^2\boldsymbol{\theta}-g \nabla\alpha \nabla\boldsymbol{\theta} )\mathcal{P}[\phi;t]-\Gamma_\theta\frac{\delta\mathcal{P}[\phi;t]}{\delta\vectheta}.
    \end{aligned}  
\end{equation}
Using the ansatz that there is a Gaussian stationary probability distribution
\begin{equation}
    \mathcal{P}[\boldsymbol{\theta},\boldsymbol{\theta}^*,\alpha] \propto \exp \left(\int_{\vecr} -\frac{Z_\theta}{2\Gamma_\theta}(\nabla\boldsymbol{\theta})^2-\frac{Z_\alpha}{2\Gamma_\alpha}(\nabla\alpha)^2\right),
\end{equation}
we are left with only  current terms which arise from our nonlinearities in the Langevin, namely 
\begin{equation}
\begin{aligned}
    J_{\text{KPZ}}^\alpha &=(\lambda_\alpha (\nabla\alpha)^2 +\lambda_\theta(\nabla\boldsymbol{\theta})^2)\mathcal{P},\\
    J_{\text{KPZ}}^{\boldsymbol{\theta}}&=g\nabla\alpha\nabla\boldsymbol{\theta}\mathcal{P}.
\end{aligned}
\end{equation}
In $d=1$ we can again identify the total derivative $\partial_i\partial^i\phi\partial_j\phi\partial^j\phi\propto\partial_i(\partial^i\phi\partial_j\phi\partial^j\phi)$. We thus require
\begin{equation}
    \int_x\frac{\lambda_\theta Z_\alpha}{\Gamma_\alpha}(\nabla\boldsymbol{\theta})^2\nabla^2\alpha+\frac{gZ_d}{\Gamma_\theta}\nabla\alpha\nabla\boldsymbol{\theta}\nabla^2\boldsymbol{\theta}=0,
\end{equation}
which holds if $\frac{\lambda_\theta Z_\alpha}{\Gamma_\alpha}=\frac{gZ_d}{\Gamma_\theta}$. Additionally $\lambda_\alpha\in\mathbb{R}$ is compatible with these conditions in $d=1$.

\section{Numerical procedures and additional results}\label{app:numerical}

\subsection{Numerical implementation}

For the numerical simulations, presented in Sec. \ref{sec:num_sim}, we have to numerically solve the stochastic first order differential equation (\ref{eq:GPE}). 
Such an equation can generally be written as 
\begin{equation}
    \frac{d\psi(t,x)}{dt}= f\left( \psi(t,x) \right) + \xi\quad,\text{with}\quad \psi,f(\cdot),\xi \in \mathbb{C}^N,
\end{equation}
where $f(\cdot)$ is the deterministic part of the equation and $\xi$ is a Gaussian white noise with zero mean, such that: 
\begin{equation}
    \langle \xi_i(t,x) \xi_j(x',t') \rangle =0 \quad \langle \xi^*_i(t,x) \xi_j(t',x') \rangle = 2 \gamma_0 \delta_{ij}\delta(x-x')\delta(t-t').
\end{equation}
We can reformulate this equation into the following form:
\begin{equation}
    d\psi(t,x)=f(\psi(t,x))dt+ dW(t),
\end{equation}
where \( W(t) \) is a stochastic process satisfying \( W(0) = 0 \), 
with normally distributed increments of zero mean and variance proportional to the elapsed time, 
and independent increments over disjoint time intervals.

This equation can be discretized in time and space, with the spacing $\Delta t$ in the time axis and $\Delta x$ in the spatial axis. 
The discretized coordinates will be named $t_i=i\Delta t$ and $x_j=j\Delta x$, with $i,j\in\mathbb{N}$. 
The spatial discretization is used to evaluate the function $f(\cdot)$, which, in our case, consists of a second derivative, that can be computed as: 
\begin{equation}
\partial^2_x\psi(t,x)\rightarrow\frac{\psi(t_i,x_{j+1})+ \psi(t_i,x_{j-1})- 2\psi(t_i,x_{j})}{ \Delta x^2},
\end{equation}
where we always used a lattice spacing of $\Delta x=1$.
To discretize the time derivative, the forward Euler-Maruyama method is used, which leads to: 
\begin{equation}
    \psi(x,t_{i+1})=\psi(x,t_{i}) + f(\psi(x,t_{i})) \Delta t+\left(W(t_{i+1})- W(t_i)\right).
\end{equation}
To implement the non-deterministic term of this expression, a random, normally distributed number with variance $\sqrt{\Gamma_0 \Delta t}$ has to be drawn for the real and imaginary part of $\Delta W$. 
This describes now a single time integration steps, which transports a state at time $t_i$ to $t_{i+1}$.
The time step is chosen as $\Delta t=0.01$ in all the simulations.

\begin{figure}[t]
    \centering
    \includegraphics[width=0.5\linewidth]{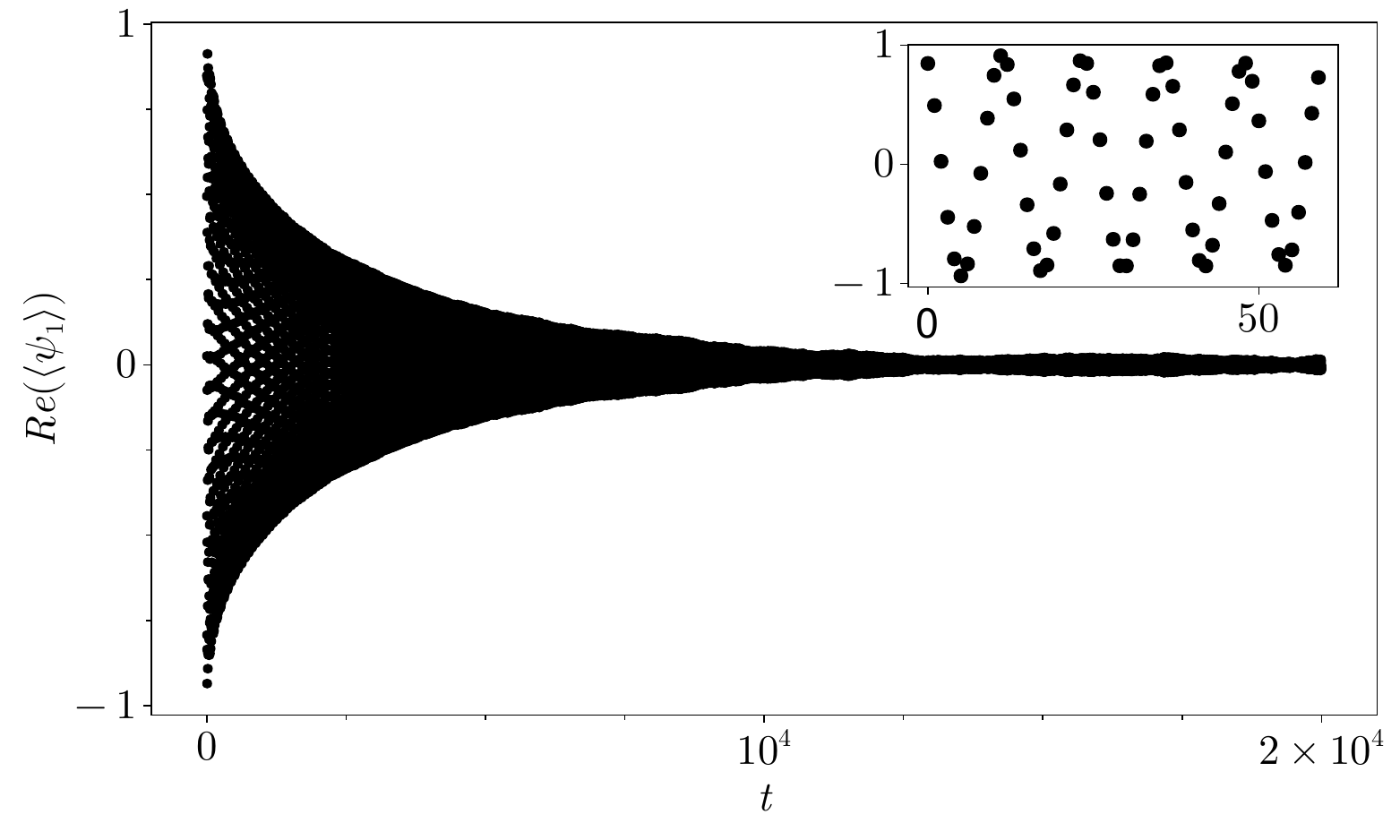}
    \caption{Real part of the expectation value of $\psi_1$ of system that is initialized as $\psi(x)=\hat{e}_1$. The couplings in the simulation are $Z_d=1$, $Z_c=-1$, $r_d=-1$, $u_d=1$, $\kappa_d=0.5$, $\kappa_c=-0.5$, $u_c=-0.2$, $r_c=-0.3$ and $\Gamma_0=0.04$ in a system with $L=1,000$ sites. The inset shows a zoom to the first 50 integration steps, where the oscillation of the field is visible.}
    \label{fig:rot_phase_N=3_gc=0.2_sigma=0.6_reaching_steady_state}
\end{figure}

To measure any observables, the system first has to be in the stationary state. 
We therefore initialized the system in the state $\psi(x)=\hat{e}_1$ and evolved in time until the expectation value $\langle \psi \rangle $ vanishes.
The temporal behavior of this expectation value is shown in Fig. \ref{fig:rot_phase_N=3_gc=0.2_sigma=0.6_reaching_steady_state}. 

To extract the scaling exponents of the correlation functions, power-law fits were performed. 
To estimate the associated uncertainties, the fitting range was systematically varied. 
More precisely, for the temporal correlation functions, the starting and ending time points of the fitting interval were varied by $\pm 20\%$, while for the spatial correlation functions the fitting range was modified by $\pm 10$ data points. 
The mean and variance of the resulting set of fitted exponents were then used to determine the final exponent values and their corresponding error estimates. 
This procedure was employed instead of the standard covariance-matrix approach, as the uncertainties obtained from the covariance matrix were several orders of magnitude smaller than those resulting from the fitting-range variation method.

\subsection{Numerical results for the other fixed points in the oscillating phase}

In the oscillating phase, the RG analysis in Sec. \ref{sec:rg} suggested, that there are two additional fixed points. 
One fixed point is stable if the three nonlinear couplings $\lambda_\alpha$, $\lambda_\theta$ and $g$ have the same sign and $N>5$.
The other fixed point is stable for every $N$, but only if $g$ has a different sign to $\lambda_\alpha$ and $\lambda_\theta$. 
We performed numerical simulations in both regimes and found no evidence of universal scaling behavior.
This absence of scaling can be attributed to the presence of topological defects, which disrupt the scaling behavior of the Goldstone modes.

\subsection{Long time limit in the oscillating phase}
\begin{figure}[t]
    \centering
    \includegraphics[width=0.5\linewidth]{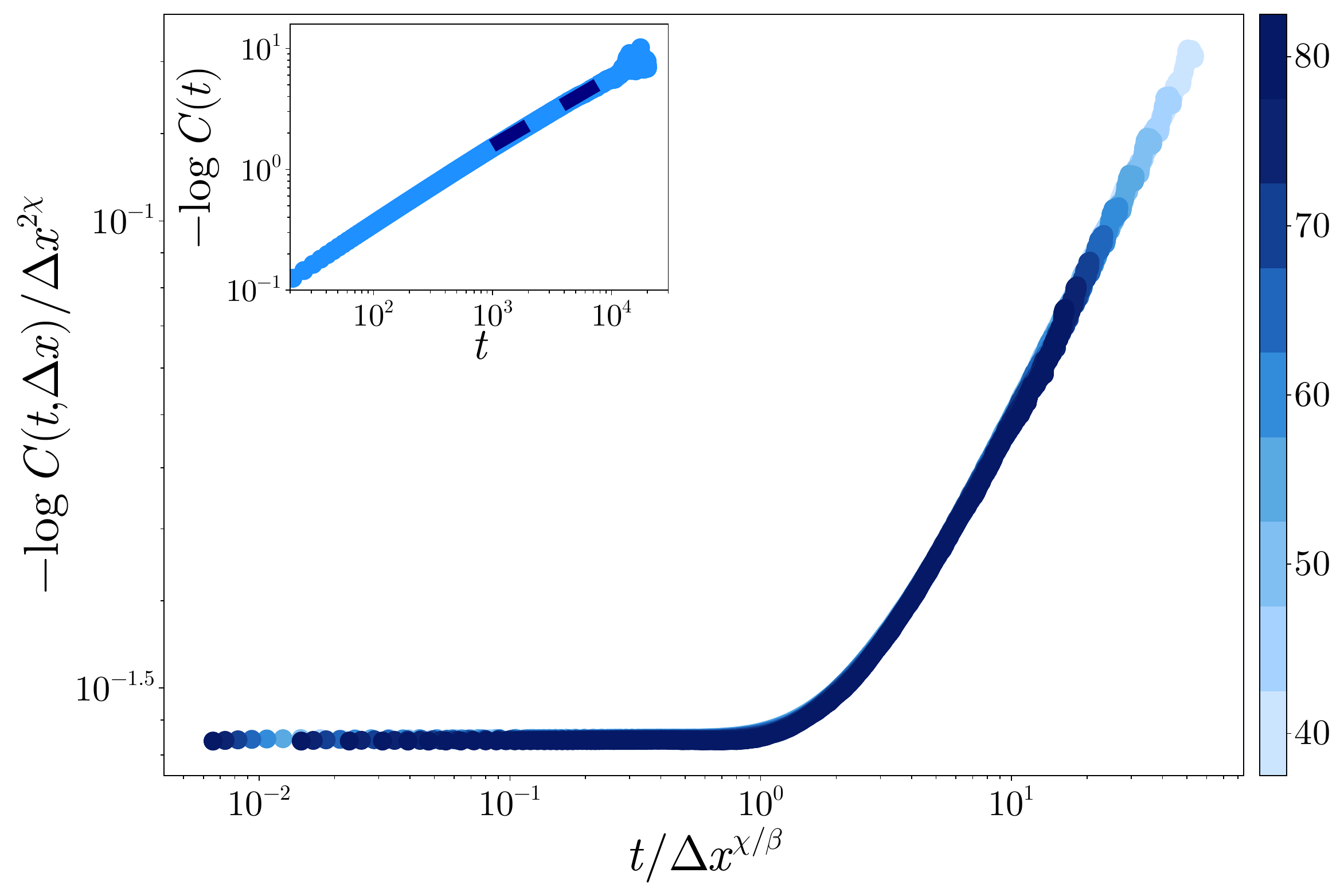}
    \caption{
    Scaling results of the functions $C_\alpha$ in the oscillating phase for $N=3$. 
    The main panel shows time correlations for spatial separations $\Delta x=45$--$80$ in steps of $5$ (lighter colors represent smaller $\Delta x$ and  darker colors represent larger $\Delta x$).
    The insert shows the time correlation with $\Delta x=0$, from which the scaling exponents $\beta_\alpha=0.279 \pm 0.004$.
    In contrast to Fig.  \ref{fig:osc_phase_N=3_KPZ}, the fit shown in the inset is performed at later times, and larger spatial separations are used for the data collapse.
    Used couplings are $Z_d=1$, $Z_c=0$, $r_d=-1$, $u_d=1$, $\kappa_d=-0.25$, $u_c=1.6$, $\kappa_c=-0.4$, $r_c=-2.1$, and $\Gamma_0=0.02$ in a system with $L=10,000$ sites. Average over 9501 paths.}
    \label{fig:osc_phase_N=3_KPZ_different_regime_where_exponents_differ}
\end{figure}

The numerical results in the oscillating phase for $N=3$ (Sec.~\ref{subsec:numOsc}) show that the scaling behavior of the correlation functions of the $\alpha$ and $\theta$ fields is consistent with the KPZ universality class.
For the $\alpha$ field, this scaling behavior is observed at intermediate time scales.
At larger time scales, however, the curve begins to flatten, as shown in the inset of Fig. \ref{fig:osc_phase_N=3_KPZ_different_regime_where_exponents_differ}.

Performing a fit in this regime yields a scaling exponent of $\beta_\alpha = 0.279 \pm 0.004$, which deviates from the KPZ value and is closer the the EW scaling exponent of $\beta_{EW}=1/4$.
In Fig. \ref{fig:osc_phase_N=3_KPZ_different_regime_where_exponents_differ}, we also show a scaling collapse using this exponent.
The collapse with this exponent occurs only for larger spatial distances compared to the KPZ collapse shown in Fig. \ref{fig:osc_phase_N=3_KPZ}.

This indicates that KPZ scaling in the oscillating phase is observable only at intermediate scales. 
The microscopic theory possesses an \(SO(2)\times O(N)\) symmetry, whose continuous components cannot exhibit conventional long-range order in one spatial dimension. 
Consequently, the description in terms of approximately independent Goldstone modes, and hence the effective scalar KPZ description of the \(\alpha\) mode, may be restricted to intermediate scales.
At larger temporal separations, the increasing importance of coupled fluctuations of the \(\alpha\) and \(\theta\) modes, finite-size effects, and possibly topological excitations can lead to deviations from KPZ scaling.
We therefore interpret the observed drift as the breakdown of the accessible KPZ scaling regime rather than as evidence for a crossover to the Edwards--Wilkinson universality class.

\subsection{Testing universality in the rotating phase}
In the rotating phase, we found weak scaling exponents that do not fit to any other known universality class. 
In this section, we additionally discuss the universality of the found scaling exponents. 
Therefore, we compare here four different simulations in which different couplings $u_c\in [-0.4,0.2]$ are chosen. 
This corresponds to 
the varying of $u_c$ in equation (\ref{eq:action}) results in varying the ratio between linear and nonlinear couplings in the effective equation of motion (\ref{eq:rot_goldstone_action}). 
For all the couplings, the system will be in the regime where defects do not dominate the scaling of the system, see Sec. \ref{sec:TopDef}.

\begin{figure}[t]
    \centering
    \includegraphics[width=0.5\linewidth]{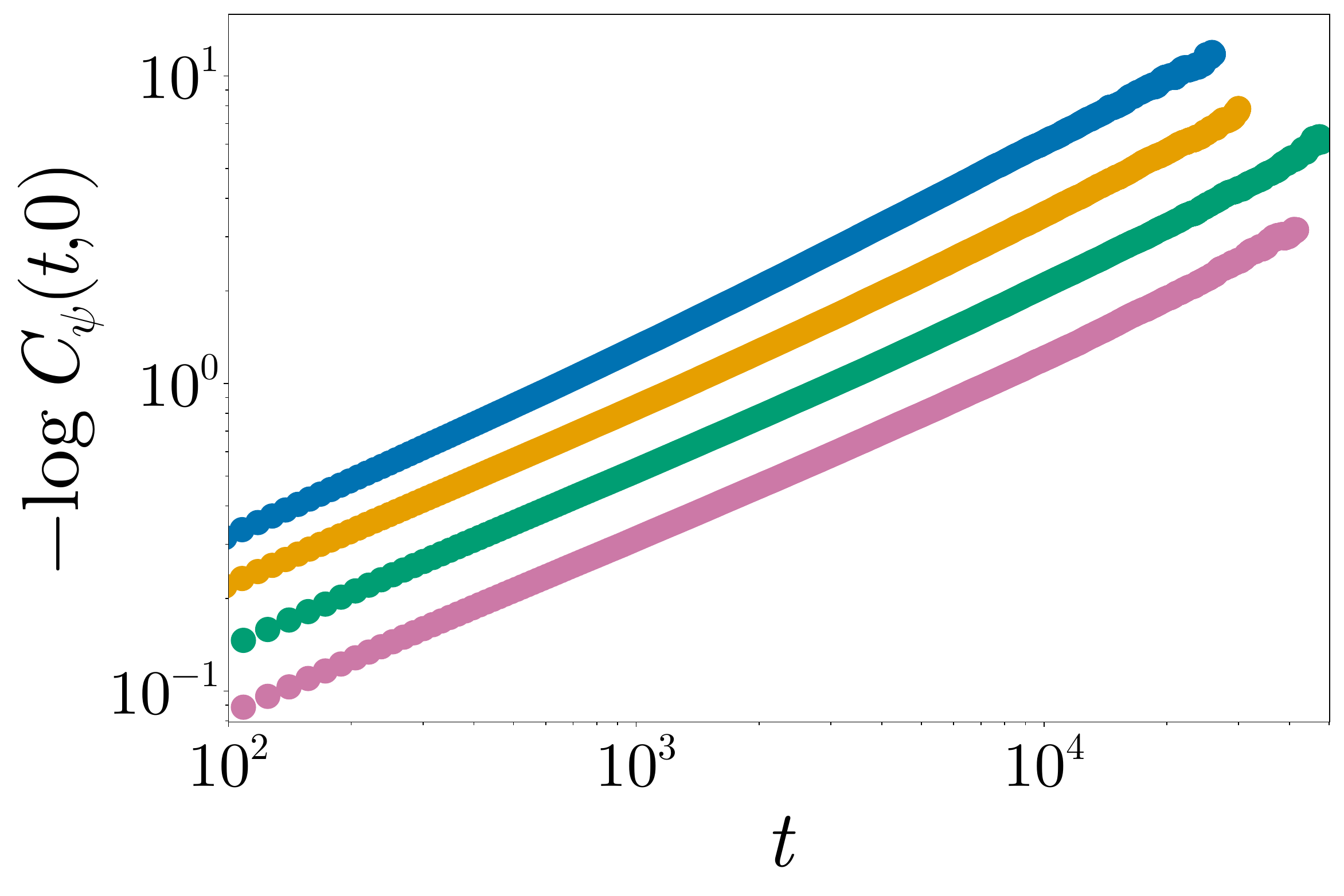}
    \caption{Time correlation $C_\psi$ in the rotating phase for different couplings. The chosen couplings are $Z_d=1$, $Z_c=-1$, $r_d=-1$, $u_d=1$, $\kappa_d=0.5$, $\kappa_c=-0.5$, $r_c=-0.3$ and $\Gamma=0.035$ in a system with $L=1,000$ sites. The coupling $u_c$ is used as a tuning parameter for the nonlinear coupling $\lambda_\alpha$.}
    \label{fig:Overview_alpha_t}
\end{figure}

\begin{figure}[t]
    \centering
    \includegraphics[width=0.5\linewidth]{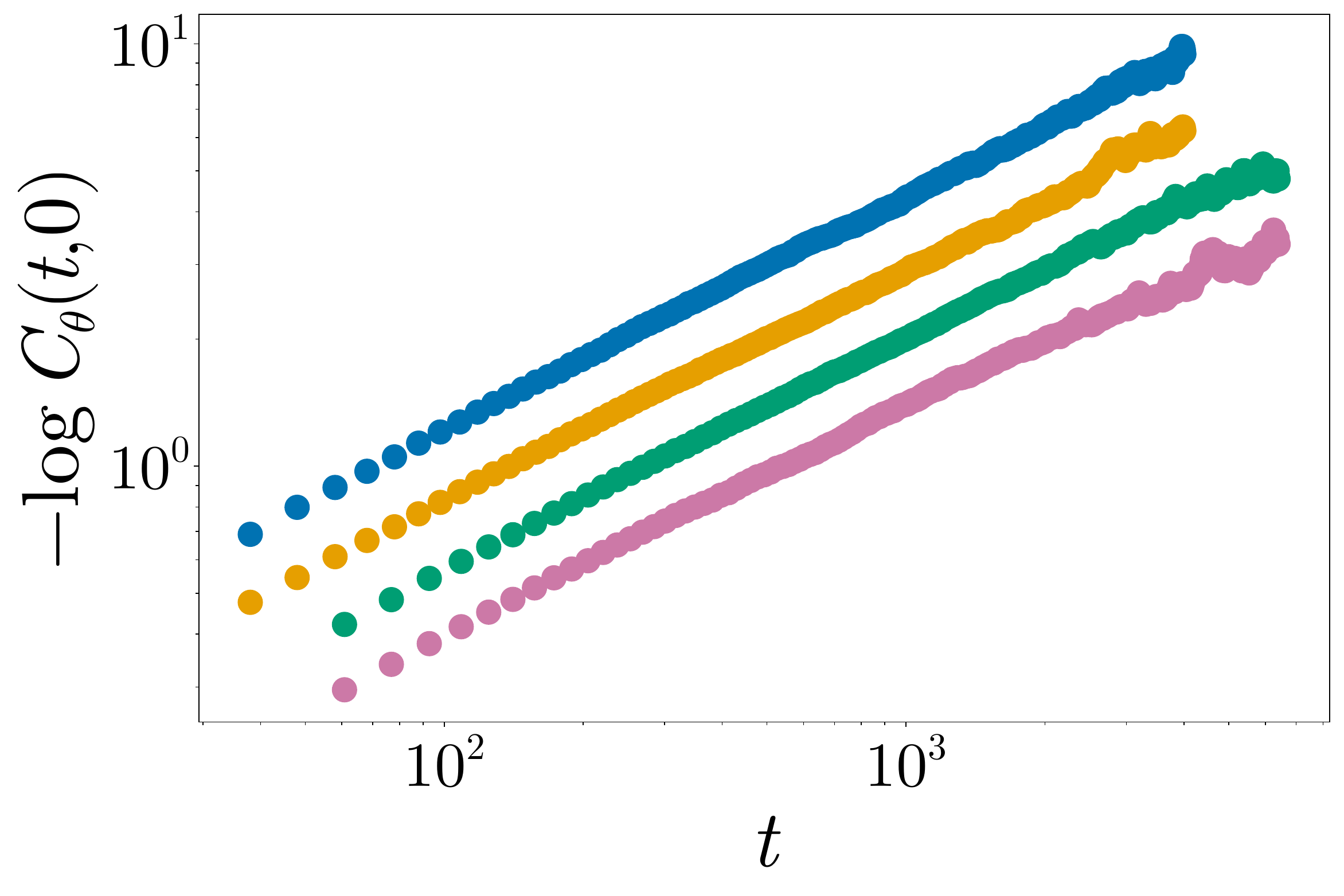}
    \caption{Time correlation $C_\theta$ in the rotating phase for different couplings. The chosen couplings the same as in Fig. \ref{fig:Overview_alpha_t} }
    \label{fig:Overview_theta_t}
\end{figure}

\begin{figure}[t]
    \centering
    \includegraphics[width=0.5\linewidth]{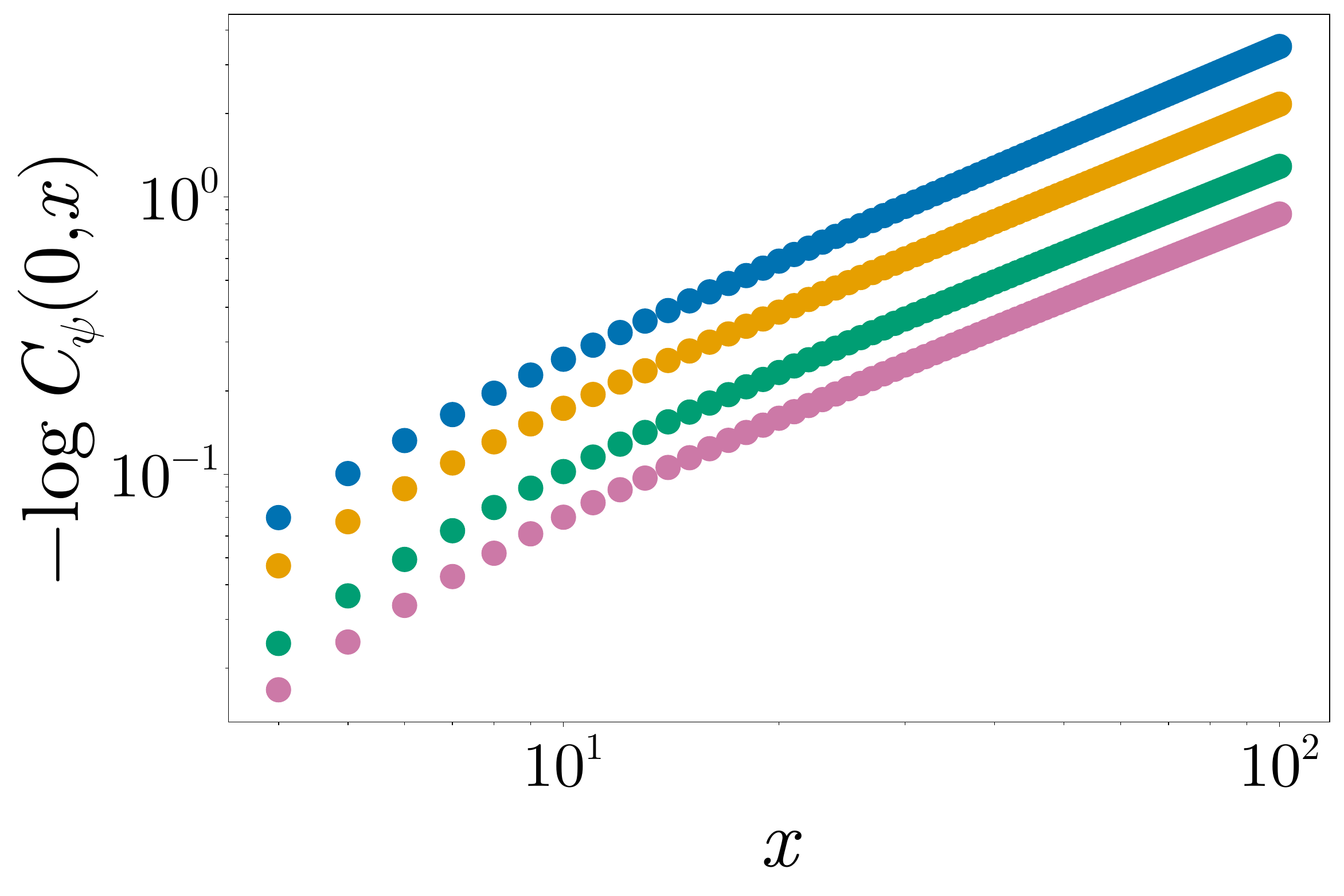}
    \caption{Spatial correlation $C_\psi$ in the rotating phase for different couplings. The chosen couplings the same as in Fig. \ref{fig:Overview_alpha_t}}
    \label{fig:Overview_alpha_x}
\end{figure}

\begin{figure}[t]
    \centering
    \includegraphics[width=0.5\linewidth]{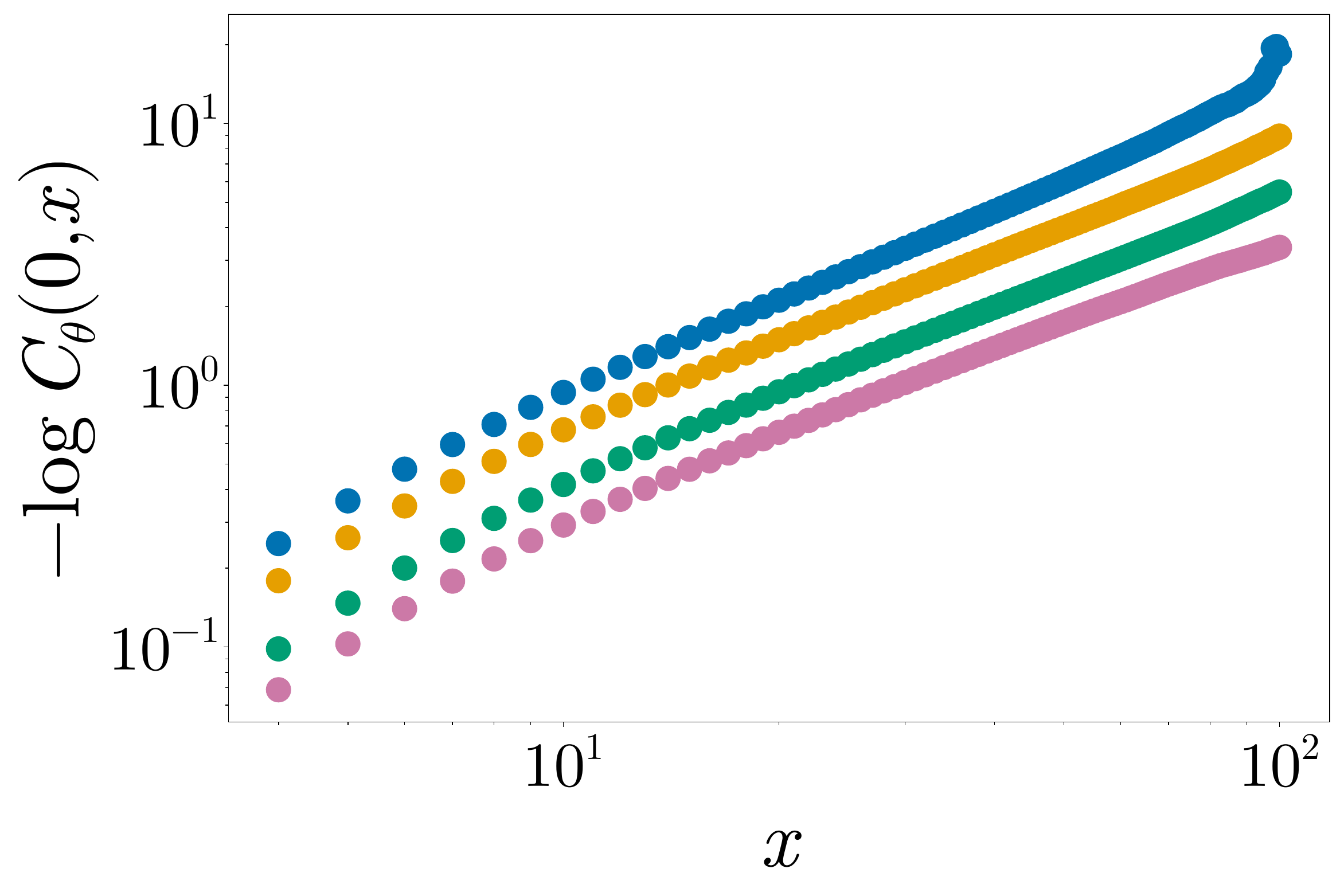}
    \caption{Spatial correlation $C_\theta$ in the rotating phase for different couplings. The chosen couplings the same as in Fig. \ref{fig:Overview_alpha_t}}
    \label{fig:Overview_theta_x}
\end{figure}

Figs.~\ref{fig:Overview_alpha_t}, \ref{fig:Overview_theta_t}, \ref{fig:Overview_alpha_x}, and \ref{fig:Overview_theta_x} show the spatial and temporal correlation functions of $C_\psi$ and $C_\theta$ for different values of $u_c$.
As illustrated in these figures, the scaling behavior of the correlation functions appears highly similar across the different coupling strengths. 
We additionally performed fits to these correlation functions to extract the corresponding scaling exponents. 
The resulting exponents for the various coupling strengths are summarized in table \ref{tab:scalingExponentsOverview_rot}. 
From this table, we observe that both the $\beta_\alpha$ and $\beta_{\theta^\sigma}$ exponents are largely independent of the coupling strength. 
In contrast, the $\chi_\alpha$ and $\chi_{\theta^\sigma}$ exponents show a slight decreasing trend with decreasing $u_c$. 
The values of the $z_\alpha$ and $z_{\theta^\sigma}$ exponents then follow from the corresponding scaling exponents.

Possible explanations for the variation of the $\chi$ exponents may be related to the presence of topological defects. 
As seen in Fig.~\ref{fig:mom_dis_rot_overview_plus_inserts}, the exponent of the momentum distribution already begins to change before the final jump occurs. 
This may indicate that a small number of topological defects already start to perturb the scaling behavior of the Goldstone modes.

Overall, the results suggest that the weak scaling behavior, as well as the values of the $\beta$ exponents, are universal within the rotating phase.



 \begin{table}
    \centering
    \begin{tabular}{|c|c|c|c|c|}
    \hline
    & $u_c=0.2$ & $u_c=0.0$ & $u_c=-0.2$ & $u_c=-0.4$ \\
    \hline
    \hline
    $\beta_\alpha$&$ 0.344(8)$& $ 0.3414(27) $ & $0.345(7)$& $ 0.343(14) $ \\
    \hline
    $\chi_\alpha$ & $0.5520(18) $ & $ 0.5339(18) $ & $0.527(4)$& $ 0.522(5) $ \\
    \hline
    $z_\alpha$ & $ 1.603(25) $ & $ 1.564(11) $ & $1.530(26)$& $ 1.52(5) $ \\
    \hline
    $\beta_{\theta^\sigma}$ & $0.2691(26)$  & $ 0.269(4) $ & $0.2706(29)$& $ 0.270(6) $ \\
    \hline
    $\chi_{\theta^\sigma}$& $ 0.571(13) $ & $ 0.537(4) $ & $0.532(4)$& $ 0.521(14) $  \\
    \hline
    $z_{\theta^\sigma}$& $ 2.12(10) $ & $ 1.99(3) $ & $1.96(3)$& $ 1.93(10) $ \\
    \hline
    \end{tabular}
    \caption{Scaling exponents of the $\alpha$ and $\theta^\sigma$ fields in the rotating phase. The fixed couplings are $Z_d=1$, $Z_c=-1$, $r_d=-1$, $u_d=1$, $\kappa_d=0.5$, $\kappa_c=-0.5$ and $r_c=-0.3$. The noise strength is $\Gamma_0 =0.035$. For The $\beta$ and $\chi$ exponents are measured through a fit and the dynamical exponents is given by the ratio $\chi/\beta$.}
    \label{tab:scalingExponentsOverview_rot}
\end{table}

\clearpage

\twocolumngrid

\bibliography{nonthermalgoldstonesMerged}

\end{document}